\documentclass[aps,amssymb,
 amsmath,
 floatfix,
 prl,twocolumn,reprint,
 tightenlines,
 longbibliography]{revtex4}
 
\usepackage{graphicx}
\usepackage{epsfig}

\usepackage{amsmath}
\usepackage{amsfonts}
\usepackage{amssymb}
\usepackage{bm}
\usepackage{mathtools}
\usepackage{comment}
\usepackage{cancel}

\usepackage[LGR,T1]{fontenc}
\usepackage[cal=boondoxo]{mathalfa}

\usepackage{hyperref}
\usepackage{dsfont}
\usepackage{lipsum}

\usepackage{tikz}
\usepackage{circuitikz}
\usetikzlibrary{arrows, shapes}
\usepackage{pgfplots}
\pgfplotsset{compat=1.14}

\usepackage{pdfpages}

\newcommand{\mean}[1]{\left\langle #1 \right\rangle}

\newtagform{supplementary}[S.]() 

\newcommand{\mainref}[1]{\begingroup\usetagform{default}\ref{#1}\endgroup}

\usepackage{comment}

\newcommand{\UNIPD}{
Department of Physics and Astronomy, University of Padova, Via Marzolo 8, I-35131 Padova, Italy}
\newcommand{\INFN}{INFN, Sezione di Padova, Via Marzolo 8, I-35131 Padova, Italy}

\begin{document}

\title{Dissipation bounds the coherence of stochastic limit cycles}
\author{Davide Santolin}
\author{Gianmaria Falasco}
\affiliation{\UNIPD}
\affiliation{\INFN}

\date{\today}

\begin{abstract}
    Overdamped stochastic systems maintained far from equilibrium can display sustained oscillations with fluctuations that decrease with the system size. The correlation time of such noisy limit cycles expressed in units of the cycle period is upper-bounded by the entropy produced per oscillation.
    We prove this constraint for first-order nonlinear systems in arbitrary dimensions perturbed by weak, uncorrelated Gaussian noise. We then extend the result to important examples of more general stochastic dynamics, including electronic and chemical clocks, illustrating the practical relevance of the dissipation-coherence bound for electronic computing and thermodynamic inference.
 
\end{abstract}

\maketitle

\emph{Introduction.} It is becoming progressively clear that fundamental constraints exist that limit the performance of physical systems operating far from thermodynamic equilibrium. Beyond the long-known Carnot bound on the efficiency of thermal engines (and its maximum-power generalization \cite{esposito2010maximum}), it has recently been discovered that dissipation limits the signal-to-noise ratio \cite{Barato2015,gingrich2016dissipation,hasegawa2019fluctuation,falasco2019unifying,horowitz2020thermodynamic,dechant2020}, the first passage times \cite{gingrich2017fundamental,falasco2020dissipation} and the response \cite{ptaszynski2024dissipation,kwon2024fluctuation,ptaszynski2024nonequilibrium} of currents, as well as the speed of transport processes \cite{shiraishi2018speed,vo2020unified,van2023thermodynamic,dieball2024thermodynamic}.
These results hold under very general conditions for mesoscopic systems subject to thermal noise and to the action of nonconservative forces. The saturation of these inequalities far from equilibrium requires optimized kinetic details that are desirable in many technological applications and, in some instances, might be selected by biological evolution \cite{marsland2019,Falasco2019a,zheng2024topological,cao2015free}.

Moreover, in the framework of Markov jump processes it has been conjectured that the entropy production $\Sigma$ (in units of $ k_B=1$) sets an upper bound on the number of coherent oscillations $\mathcal{N}$ an overdamped stochastic system can exhibit \cite{oberreiter2022universal}. Specifically, $\mathcal{N}$, defined as the product of the correlation time $\tau_c$ and the mean oscillation frequency $\omega=2\pi/t_p$, should satisfy $\mathcal{N}:= \omega \tau_c   \leq  \Sigma/ (2\pi)$. 
The inequality---yet unproven despite recent progress
\cite{ohga2023thermodynamic,shiraishi2023entropy}---may be relevant in a variety of systems that utilize autonomous oscillations to synchronize multiple mesoscopic processes, such as circadian rhythm in bacteria \cite{Monti2018robustness}, calcium spiking in mitochondrial metabolism \cite{voorsluijs2024calcium}, and clock signals in microprocessors \cite{dadalt2018understanding}.

Here we prove this bound for the overdamped Langevin dynamics
    \begin{align}\label{Langevin}
        \dot x(t) = F(x(t)) + \sqrt{2\epsilon} \xi(t),
    \end{align}
where $x \in \mathbb{R}^d$, $\xi$ is a weak ($\epsilon \to 0$), zero-mean Gaussian noise with covariance $\mean{\xi(t)\xi(0)}=  \delta(t) D(t)  $, and $F$ is a nonconservative drift field.

\begin{figure*}[htbp]    
        \includegraphics{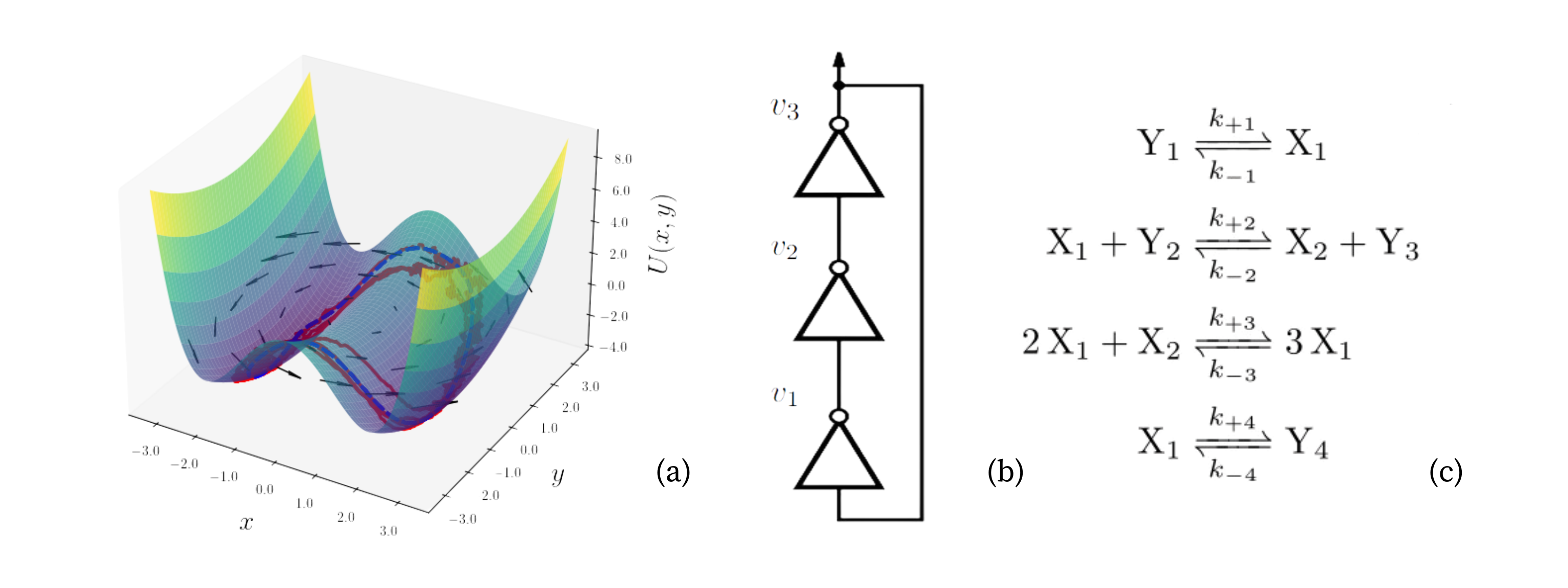}
        \caption{The coherence-dissipation bounds, Eqs. \eqref{series_bounds} and \eqref{bound_micro}, apply to a wealth of systems ranging from stirred colloidal particles in bistable potentials (a), to electronic circuits and chemical reaction networks, such the ring oscillator (b) and the autocatalytic Brusselator model (c). In panel (a), the black arrows represent the rotational flow and the solid red line shows the noisy limit cycle, which reduces to the deterministic attractor (dashed blue line) in the noiseless limit.}
        \label{fig: models}
    \end{figure*}

Therefore, the coherence-dissipation bound, Eq. \eqref{series_bounds}, holds for a wealth of soft-matter physics systems, such as driven colloids or polymers, when the typical energy is much larger than the thermal energy scale set by an equilibrium environment. 
Equation \eqref{Langevin} also emerges as an effective description valid in the thermodynamic limit for systems described at microscopic level as interacting Markov jump processes \cite{RevModPhys.97.015002}, such as electronic circuits \cite{Freitas2020linear,Freitas2020nonlinear,gopal2022large,gopal2024thermodynamic} and chemical reaction networks \cite{vanKampen,gillespie2001approximate,lazarescu2019large,boland2008limit} (see Fig. \ref{fig: models}). 

We provide an analytical proof of the coherence-dissipation bound for the latter systems in the vicinity of a Hopf bifurcation, Eq. \eqref{bound_micro}, and demonstrate numerically that it remains valid far from this critical point.  

\textit{Weak-noise analysis.} We focus on systems that display (at least) one attractive limit cycle, i.e., \eqref{Langevin} admits for $\epsilon=0$ a stable $t_p$-periodic orbit $\mathcal{x} (t)=\mathcal{x} (t+t_p)$ of length $\ell$. For $0< \epsilon \ll 1 $, thermal noise leads the quasiperiodic stochastic paths to decorrelate exponentially as $\mean{x(t)x(0)} \underset{\epsilon \to 0}{\sim} e^{-t/\tau_c}$.
The correlation time $\tau_c$, which diverges as $\epsilon \to 0$, can be obtained from a WKB expansion of the Fokker-Planck equation associated with \eqref{Langevin} \cite{graham1987macroscopic}. Using the asymptotic form of the probability density $p(x,t)\underset{\epsilon \to 0}{\sim} e^{-I(x,t)/\epsilon} $, and approximating the rate function $I(x,t)$ with a parabola, one reduces \eqref{Langevin} to the deterministic, linear dynamics $
\dot x= J\cdot x + 2 D \cdot \pi$ for the most probable paths \cite{vance1996fluctuations,gaspard2002clocks}. Here, $J(t)$ is the Jacobian matrix of $F$ evaluated on the limit cycle $\mathcal{x} (t)$ and $\pi(t)$ describes the typical noise realizations, obeying the equation \cite{dykman1994large}
    \begin{align}\label{linear}
        \dot \pi= - J^\mathrm{T} \cdot \pi.
    \end{align}
Floquet theory leads to the correlation time \cite{vance1996fluctuations,gaspard2002trace}
    \begin{equation}
        \frac{1}{\tau_c} = \frac{(2 \pi)^2 \epsilon}{t_p^3}\int_0^{t_p} dt\, \zeta \cdot D \cdot \zeta,
        \label{eq_VanceBeta}
    \end{equation}
where $\zeta$ is the leading Floquet mode of \eqref{linear}. Since the leading Floquet mode of the deterministic linearized dynamics $\dot x= J\cdot x$ (the adjoint of \eqref{linear}) is $F(\mathcal{x} (t))$, the orthonormality condition imposes for all $t$ 
    \begin{align}\label{eq:biorthonormal}
        \zeta(t) \cdot F(\mathcal{x}(t) )=1.
    \end{align}
In particular, $\zeta=1/F(\mathcal{x} )$ for $d=1$, which allows us to express \eqref{eq_VanceBeta} in a closed form.

To leverage the explicit knowledge of the correlation time in 1-dimensional systems we project \eqref{Langevin} onto the direction tangent to the limit cycle, identified by the unit vector $u (t):=F(\mathcal{x} (t))/|F(\mathcal{x} (t))|$. The dynamics of the tangent coordinate $x_\parallel := u \cdot x $ obeys
    \begin{align}\label{Langevin_parallel}
        \dot x_\parallel = |F| + g(x_\parallel,x_\perp)+ \sqrt{2D_\parallel\epsilon} \eta(t)
    \end{align}
with diffusion constant $D_\parallel =  D : uu$, $\eta$ a zero-mean Gaussian white noise with unit variance, and $g$ a function depending on the transverse fluctuations,  $\mathbb{R}^{d-1} \ni x_\perp := (\mathbb{I}- u u )\cdot x $. 
In what follows, we need not examine the exact dynamics \eqref{Langevin_parallel}, but it suffices to consider the auxiliary dynamics obtained by setting $g$ to zero in \eqref{Langevin_parallel}, i.e., decoupling the tangent and perpendicular motion (see SM \footnote{SM at [URL inserted by the publisher] include the details of the analytical and numerical calculations.}, Sect. I). The associated  mean entropy production rate reads \footnote{We ignore the variation of the system entropy as it  vanishes in the macroscopic limit \cite{RevModPhys.97.015002}.}
    \begin{align}\label{epr_auxiliary}
        \dot \Sigma^{(1)} = \mean{\frac{|F| \dot x_\parallel}{ \epsilon D_\parallel }}  \underset{\epsilon \to 0}{=} \frac{|F|^2}{  \epsilon D_\parallel}\bigg|_{x=\mathcal{x}(t) }  \geq 0,
    \end{align}
where we have used that the system probability peaks on the deterministic trajectory when $\epsilon \to 0$ \cite{RevModPhys.97.015002}.

Equation \eqref{eq_VanceBeta} specialized to the auxiliary dynamics yields the correlation time \cite{Remlein2022coherence}
    \begin{align} \label{tc_1D}
            \tau^{(1)}_c&= \frac{1}{(2 \pi)^2 \epsilon } \dfrac{t_p^3}{\int_0^{t_p} dt D_{\parallel}/|F|^2 }.
    \end{align}
By using \eqref{epr_auxiliary} and exploiting the inequality between the harmonic and the arithmetic mean (see SM, Sect. II) we derive that the number of coherent oscillations of the auxiliary dynamics $\mathcal{N}^{(1)} :=2 \pi \tau^{(1)}_c/t_p$ is upper bounded by the corresponding entropy production integrated along the cycle, $\Sigma^{(1)} :=\int_0^{t_p} dt \dot \Sigma^{(1)} $:
    \begin{equation}\label{bound_N1}
         \mathcal{N}^{(1)} = \dfrac{t_p^2}{2\pi} \left( \int_0^{t_p}  \frac{dt}{\dot \Sigma^{(1)}}\right)^{-1}   \leq \frac{ \Sigma^{(1)}}{2 \pi  }.
    \end{equation}
In the following, we wish to weaken the bound \eqref{bound_N1} (first appearing in \cite{remlein2024nonlinear}) to relate $\mathcal{N}^{(1)}$ and $\Sigma^{(1)}$ to the physical quantities $\mathcal{N}$ and $\Sigma$, respectively.

First, it follows from the Cauchy-Schwarz inequality (see SM, Sect. III) that $\dot \Sigma^{(1)}$ is never larger than the entropy production rate associated to the $d$-dimensional Langevin dynamics \eqref{Langevin},
    \begin{align}\label{epr_D}
        \dot \Sigma (t)   \underset{\epsilon \to 0}{=} \frac{1}{\epsilon} F\cdot D^{-1} \cdot F|_{x=\mathcal{x }(t)} \geq \dot \Sigma^{(1)}(t).
    \end{align}
In particular, \eqref{epr_D} holds as an equality when the diffusion matrix $D$ is proportional to the identity.
Second, notice that $\mathcal{N}^{(1)}=\mathcal{N}$ if the tangent dynamics \eqref{Langevin_parallel} is decoupled from orthogonal fluctuations, that is, if $g \equiv 0$. In general, we complete the proof by showing that 
    \begin{align}\label{bound_N_N1}
        \mathcal{N}^{(1)} \geq \mathcal{N} \quad \quad \text{ if } D(t) = c(t)\mathbb{I},
    \end{align}
for any positive function $c$. For arbitrary $d>1$ we split (the generally  unknown) $\zeta$ in the components tangent and perpendicular to the limit cycle, i.e. $\zeta:= \zeta_\parallel  +\zeta_\perp $ with $\zeta_\parallel  \cdot \zeta_\perp=0$, $\zeta_\parallel= F(\mathcal{x} )/|F(\mathcal{x} )|^2$ and $\zeta_\perp \cdot F(\mathcal{x} ) =0$, such that \eqref{eq:biorthonormal} is satisfied. Hence, we can write \eqref{eq_VanceBeta} as
    \begin{equation}\label{bound_tau_c}
         \frac{1}{\tau_c} \geq  \frac{1}{\tau^{(1)}_c} + \frac{8 \pi^2 \epsilon}{t_p^3} \int_0^{t_p} dt\, \zeta_\parallel \cdot D \cdot \zeta_\perp,
    \end{equation}
dropping the positive contribution $\int_0^{t_p} dt\, \zeta_\perp \cdot D \cdot \zeta_\perp \geq 0$ that stems from fluctuations transverse to the limit cycle.
If $D$ is proportional to the identity matrix, the integrand in the last term of \eqref{bound_tau_c} is null, implying $\tau^{(1)}_c \geq \tau_c $ and thus \eqref{bound_N_N1}. Collecting the results \eqref{bound_N1}, \eqref{epr_D},  and \eqref{bound_N_N1} we arrive at the first main result of the Letter,
    \begin{equation}\label{series_bounds}
         \mathcal{N} \leq       \mathcal{N}^{(1)} \leq \frac{\Sigma^{(1)}}{2 \pi  }  \leq \frac{\Sigma}{2 \pi  } ,
    \end{equation}
which is the conjectured bound. In particular, the normal form of the Hopf bifurcation can be shown to saturate \eqref{series_bounds} (see SM, Sect. IV). 

\textit{Stirred colloidal particle.} We exemplify \eqref{series_bounds} by considering the $d=2$ motion of a colloidal particle in a solvent at temperature $T$ and bistable potential $U(x,y)=\alpha x^4 - \beta x^2 + \delta y^2$, with $\beta >0$, forced by a nonconservative rotational flow $f = \gamma (-y, x)^T$, i.e., $F= \mu(-\nabla U +f)$, with $\mu$ the particle mobility, see Fig. \ref{fig: models}a. The scale provided by the energy barrier $\Delta U$ of the double-well potential allows one to rescale energies as $ T/\Delta U=\epsilon$, this quantity being small $\epsilon\ll 1$ for comparatively large barriers. With this scaling and ignoring hydrodynamic boundary effects, the diffusion matrix reads $D = \mathbb{I}$ with the appropriate time units to set $\mu=1$.

\begin{figure}[t]
        \includegraphics[width=\linewidth]{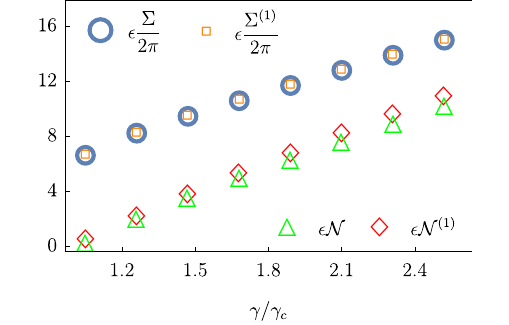}
        \caption{Numerical evaluation the quantities in inequalities \eqref{series_bounds} for a colloidal particle in a bistable potential as a function of the rotational flow strength $\gamma$ normalized to the bifurcation value $\gamma_c \simeq 1.19$., with parameters $\alpha=0.25,\beta=2$ and $\delta=0.25$.}
        \label{fig: hierarchyDW}
    \end{figure}

A critical value of the driving strength $\gamma$ exists, at which the dynamics \eqref{Langevin} for $\epsilon=0$ undergoes a global bifurcation. At this point, two symmetric stable fixed points coalesce into a stable limit cycle, illustrated in Fig. \ref{fig: models} (a). Figure \ref{fig: hierarchyDW} shows the predicted hierarchy of bounds given by \eqref{series_bounds}, demonstrating that the theory outlined above holds for generic periodic attractors, regardless of the type of bifurcation and its proximity.

The observed scaling with the driving strength $\gamma$ can be explained analytically as follows. Solving \eqref{Langevin} for $x_2$ in $\epsilon =0$, we find that $x_1$ behaves for $\gamma \gg 1$  as a Van der Pol oscillator with weak damping (see SM, Sect. V), which is known to have oscillations with amplitude $\ell \sim \gamma^0$ \cite{sanders2007averaging}. This mapping implies that the period of the oscillations scales asymptotically as $ t_p \sim \ell/|F| \sim \gamma^{-1}$, since $|\dot x| = |F| \sim \gamma$ for $\gamma \gg 1$.  Then, it is straightforward to obtain $\tau_c^{(1)}\sim \gamma^0$ from \eqref{tc_1D}, which leads to $\mathcal{N}^{(1)} \sim \tau_c^{(1)}/t_p\sim\gamma$. Analogously, $\mathcal{N}\sim\gamma$ because it follows from Eq. \eqref{eq:biorthonormal}  that $\zeta\sim\gamma^{-1}$ and so $\tau_c\sim\gamma^0$ also. Here, we approximate integrals over one period by the leading order of the integrand times the duration $t_p$. The linear scaling of the drift field together with Eqs. \eqref{epr_auxiliary} and \eqref{epr_D}  also imply that both $\dot \Sigma$ and $ \dot \Sigma^{(1)}$ grow as $ \sim \gamma^2$. From these considerations, we also obtain a linear growth for $\Sigma$ and $\Sigma^{(1)}$, since $\Sigma \sim t_p\dot \Sigma \sim \gamma$. Therefore, the example shows that \eqref{series_bounds} can remain tight even far from equilibrium.

\emph{Extension to Markov jump processes.} We can generalize \eqref{series_bounds} to interacting Markov jump processes that possess a macroscopic limit, e.g., chemical reaction networks, electronic devices, driven mean-field Potts models \cite{RevModPhys.97.015002}. Their small Gaussian fluctuations are described by \eqref{Langevin} linearized around the deterministic attractor $\mathcal{x} (t)$, according to Van Kampen's system size expansion \cite{vanKampen}. The drift field $F= \sum_\rho \Delta_\rho j_\rho$ is constructed from the microscopic flux $j_\rho(x)$ of the transition $\rho$ (with jump size $\Delta_\rho $), while the diffusion matrix reads $D(t)= \sum_\rho \Delta_\rho \Delta_\rho j_\rho(\mathcal{x}(t))/2$. To extend the proof of \eqref{series_bounds}, we rely on the fact that the microscopic entropy production rate defined through the master equation as
    \begin{align}\label{eq MEentropy}
        \dot \Sigma^{\text{ME}}  \underset{\epsilon \to 0}{=} \frac{1}{\epsilon} \sum_{\rho>0} (j_\rho-j_{-\rho}) \ln (j_\rho/j_{-\rho}) \geq \dot \Sigma,
    \end{align}
is underestimated by $\dot \Sigma$, which is merely a measure of the dynamical irreversibility of Langevin trajectories \cite{RevModPhys.97.015002}. For instance, it is intuitively clear that  $\dot \Sigma$ misses the housekeeping entropy production of the microscopic jump dynamics: $\dot \Sigma \underset{\epsilon \to 0}{=} 0$ in any fixed point $x^*$ because $F(x^*)=0$, while $\dot \Sigma^{\text{ME}}$ is positive if the dynamics lacks detailed balance. Therefore, conditions \eqref{eq MEentropy} and \eqref{epr_D} yield  $ \Sigma^{(1)} \leq \Sigma \leq  \Sigma^{\text{ME}}$ for the period-integrated quantities.

Furthermore, one needs to account for a diffusion matrix $D$ that is not proportional to the identity, or even nondiagonal as microscopic transitions $\rho$ can simultaneously change different elements of the state vector $x$. For general positive-definite $D$, the last term in \eqref{bound_tau_c} need not be null or positive. Hence, for $\tau^{(1)}_c \geq \tau_c $ and \eqref{series_bounds} to hold, the condition $ \mathcal{I}:=\int_0^{t_p} dt\, (\zeta_\perp \cdot D \cdot \zeta_\perp + 2\zeta_\parallel \cdot D \cdot \zeta_\perp  ) \geq 0$ must be satisfied.
Hereafter, we show numerically that it is indeed the case for paradigmatic models of electronic and chemical clocks, and prove it analytically for small-amplitude oscillations close to a supercritical Hopf bifurcation. Under these conditions we arrive at the second main result of the Letter
    \begin{align}\label{bound_micro}
        \mathcal{N} \leq \frac{\Sigma}{2\pi  } \leq \frac{\Sigma^{\text{ME}}}{2\pi  }.
    \end{align}

    \begin{figure}[t]
        \includegraphics[width=\linewidth]{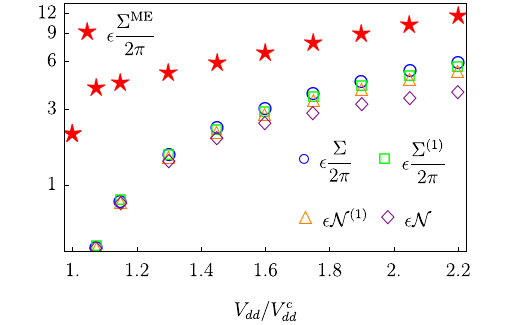}
        \caption{Numerical evaluation of the quantities in \eqref{series_bounds} for the ring oscillator with $N=3$ inverters as a function of the driving voltage $V_{dd}$ normalized by the bifurcation value $V_{dd}^c/V_T= \ln 3$. }
        \label{fig:RObound}
    \end{figure}

\emph{Ring oscillator.} An example of a diagonal $D$ not proportional to the identity matrix appears in the ring oscillator \cite{Hajimiri1999}, the archetype of electronic clocks. 
This device consists of an odd number, $N$, of CMOS inverters connected in series, where the output current of the final inverter is fed back as input to the first, each subjected to a voltage bias $V_{dd}$ (see Fig. \ref{fig: models} (b)). 
For sufficiently large voltages $V_{dd}$ the macroscopic currents show sustained oscillations.  For the potential $x_i$ after the inverter $i$, the drift has the form $ F_i =[I_p(x_{i}, x_{i-1}) - I_n(x_{i}, x_{i-1})]/C $,    with $x_i$ (resp. $x_{i-1}$) being the output (resp. input) potential at the inverter $i=1,\dots,N$ (modulo $N$) and $C$ the inverter equivalent capacitance. $I_p$ and  $I_n$ are the two MOS currents flowing into each inverter, reading in the subthreshold operation regime \cite{gopal2022large}
    \begin{align}
        I_p(x_{out}, x_{in}) &= \dfrac{C}{\Gamma}\mathrm{exp}\left(\dfrac{V_{dd}-x_{in}}{V_T}\right)\nonumber\\ 
        &\times\left[ 
        1- \mathrm{exp}\left( -\dfrac{V_{dd}-x_{out}}{V_T} \right) \right],
        \label{eq Subthreshold CMOS} 
    \end{align}
and $I_n(x_{out}, x_{in}) = I_p(-x_{out}, -x_{in})$, with $\Gamma$ the inverter characteristic time, which we set to $\Gamma=1$ in the following. In addition, we will scale all voltages by the thermal voltage $V_T=T/q_e$, with $q_e$ the electron charge, and work with the dimensionless variable $v_i :=x_i/V_T$.
The only fixed point of the dynamics \eqref{Langevin} in $\epsilon=0$ is $v^* =0$, characterized by a supercritical Hopf bifurcation at $V_{dd}^c=V_T \ln \left[1+\sec(\pi/N)\right]$ \cite{gopal2024when}. Due to the translational invariance of the ring oscillator, the stable limit cycle observed for $V_{dd}>V_{dd}^c$ obeys the relation $ \mathcal{v}_i(t) = \mathcal{v}_{i+1}(t+\Delta t)$, where $\mathcal{v}(t)=\mathcal{v}(t+t_p)$ is the deterministic periodic solution and $\Delta t$ is a delay time set by the inverter capacitance. 

The expression \eqref{eq Subthreshold CMOS} is the macroscopic limit of the microscopic discrete flow of elementary charges $q_e$ that jump across conductive channels with bi-Poisson statistics \cite{Freitas2020nonlinear}. This limit follows from the fact that transition rates are proportional to the capacitance $C$ which in turn scales as the system size, making the effect of the single charge transition on the voltages (i.e. $\pm q_e/C$) infinitesimal \cite{Freitas2021linear,RevModPhys.97.015002}.
Hence, Eq. \eqref{Langevin} for the ring oscillator results from a weak-noise approximation of the master equation, in which the diffusion matrix is diagonal with elements 
    \begin{align}
        & D_{ii}(t) = \mathrm{e}^{V_{dd}/V_T}\mathrm{cosh}(\mathcal{v}_i(t)) + \mathrm{cosh}(\mathcal{v}_i(t)-\mathcal{v}_i(t+\Delta t))
            \label{eq DiffusionElement}.
    \end{align}

Thanks to the diagonal form of $D$ we can prove that $\mathcal{I>0}$ is satisfied close to the bifurcation point.
We expand the small-amplitude limit cycle $\mathcal{v}_i(t) \sim \varepsilon $ around the fixed point $v^* =0$, with $\varepsilon :=\sqrt{V_{dd}-V_{dd}^c} \ll 1$ \cite{gopal2024when}, and the Floquet mode as
$\zeta_{\alpha} = \zeta_{\alpha}^{(0)} + \varepsilon\zeta_{\alpha}^{(1)} + \mathcal{O}(\varepsilon^2)$ ($\alpha= \parallel, \perp$) with $|\zeta_{\alpha}^{(0)}| \sim \varepsilon^0 \sim |\zeta_{\alpha}^{(1)}| $.
Likewise expanding the biorthonormality condition \eqref{eq:biorthonormal},
we obtain $\zeta_{\perp}^{(0)} \cdot \zeta_{\parallel}^{(0)} = 0$ and $\zeta_{\perp}^{(0)} \cdot\zeta_{\parallel}^{(1)} + \zeta_{\perp}^{(1)} \cdot\zeta_{\parallel}^{(0)} = 0$ for the first two leading orders. Plugging these conditions into $ D_{ii}(t) = \overline{D} + \mathcal{O}(\varepsilon^2) \label{eq Dexpanded}$,  with the fix-point value $\overline{D} =  1 + \mathrm{e}^{V_{dd}/V_T}$, leads to $\mathcal{I} =  \overline{D}\int_0^{t_p} |\zeta_{\perp}^{(0)}|^2 + \mathcal{O}(\varepsilon^2)$. Since $t_p \sim \varepsilon^0$ close to the Hopf bifurcation, we obtain that $\mathcal{I}>0$ is satisfied at leading order in $\varepsilon$. Numerical simulations of the simplest realization of the ring oscillator, which involves $N=3$ inverters (see SM, Sect. VI), show that \eqref{bound_micro} is verified even far from the bifurcation, see Fig. \ref{fig:RObound}. 

\emph{Validity of the bound close to Hopf bifurcations.} While the analysis of the ring oscillation is based on the specific form of the diffusion matrix at the bifurcation, it can be extended to any system \eqref{Langevin} with nondiagonal $D(t)$. Using the change of variable $z(t)=A(t)^{-1}(x(t)-x^*_c)$, with $ A=D^{1/2}$ and $x^*_c$ the fixed point at the onset of the supercritical Hopf bifurcation, one transforms \eqref{Langevin} into a Langevin equation with uncorrelated noise components. In view of the above proof, inequalities \eqref{series_bounds} hold for the number of coherent cycles of $z(t)$, $\mathcal{N}^{(z)}$, and its associated entropy production, $\Sigma^{(z)}$, as $\mathcal{N}^{(z)} \leq \Sigma^{(z)}/(2 \pi)$. Since it can be shown (see SM, Sect. VII) that $\mathcal{N} \leq \mathcal{N}^{(z)}$ and $\Sigma^{(z)}  \leq \Sigma$ as $\epsilon \to 0$ and at leading order in the bifurcation distance $|\mathcal{x} -x^*_c| \ll 1$, one concludes that \eqref{bound_micro} is valid close to any supercritical Hopf bifurcation irrespective of the specific form of $D$. 

\emph{The Brusselator.} A paradigmatic example of chemical oscillations is provided by the dynamics of autocatalytic reaction networks \cite{blokhuis2020universal,wilhelm2009smallest}. In particular, the Brusselator model \cite{Prigogine1968} is a minimal system that sustains limit cycles and spatio-temporal patterns far from equilibrium \cite{nguyen2018phase,Falasco2018Turing,Avanzini2019waves}. This model consists of four reactions involving two dynamical chemical species with concentrations $x_i$ and four chemostated species with fixed concentration $y_i$, see Fig. \ref{fig: models} (c).
When the chemical potentials $\mu_i = T\ln y_i$ of the chemostated species are different, two thermodynamic forces appear, $\mathcal{A}_a=\ln (y_1/y_4)$ and $\mathcal{A}_b=\ln (y_2/y_3)$, that break the detailed balance condition. If the reactions follow mass-action kinetics in a well-stirred solvent at temperature $T$ where a large number of molecules are present, a weak-noise approximation maps the microscopic jump dynamics onto \eqref{Langevin}. The associated drift field and diffusion matrix (see SM, Sect. VIII) are built from the fluxes $j_1=k_1 y_1$, $j_{-1}=k_{-1}x_1$, $j_2 = k_2x_1y_2$, $j_{-2} = k_{-2}x_2 y_3$, $j_3 = k_3x_1x_2$, $j_{-3} = k_{-3}x_1^3$, $j_4 = k_4 x_1$, $j_{-4} = k_{-4} y_4$ and the jump vectors $\Delta_{\rho}^1 = (1, -1, 1, -1)$, $\Delta_{\rho}^{2} = (0, 1, -1, 0)$.

\begin{figure}[t]
        \includegraphics[width=\linewidth]{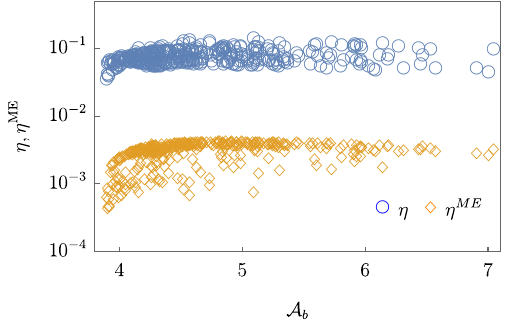}
    \caption{The ratios $\eta$ and $\eta^{\text{ME}}$ defined in the main text for the Brusselator model with $k_1 y_1 = 1.4$, $k_{-1} = k_2 = k_{-2} = k_3 = k_{-3} = k_4 =1$, $k_{-4}y_4= 0.01$, obtained numerically by sampling uniformly $y_2 \in [3, 9.5]$ and $y_3 \in [10^{-4}, 0.13]$ and restricting to 385 values that result in an oscillatory steady state.}
    \label{fig:brusBound}
\end{figure}

The tightness of the bounds \eqref{bound_micro} is encoded in the ratios $\eta^{\text{ME}} :=    2\pi \mathcal{N}/\Sigma^{\text{ME}} \leq   \eta :=2\pi \mathcal{N}/\Sigma \leq 1$, which we evaluate on different limit cycles obtained by randomizing $\mathcal{A}_b$ and keeping $\mathcal{A}_a$ fixed. As seen in Fig. \ref{fig:brusBound}, \eqref{bound_micro} are always satisfied regardless of the distance from the bifurcation, as already observed in the ring oscillator. The microscopic entropy production $\Sigma^{\text{ME}}$ far exceeds its coarse-grained version $\Sigma$, though, resulting in a loose bound. No qualitative differences are observed if the role of $\mathcal{A}_b$ and $\mathcal{A}_a$ is exchanged (see SM, Sect. VIII).

\emph{Conclusion.} In this Letter, we have proven that dissipation bounds the number of coherent oscillations, Eq. \eqref{series_bounds}, in overdamped Langevin systems with weak Gaussian noise. While the result is generic for uncorrelated noise, we also proved \eqref{bound_micro} for the case of correlated noise components, as emerging in the macroscopic limit of jump processes, close to the onset of oscillations. Moreover, we numerically showed the validity of the bound \eqref{bound_micro} arbitrarily far from the bifurcation in models of electric and chemical oscillators.
The result has profound relevance for electronic computation \cite{kish2002end,gao2021principles,Freitas2021reliability,konopik2023fundamental,wolpert2024stochastic} and thermodynamic inference in biology \cite{li2019quantifying,seifert2019stochastic,skinner2021estimating,harunari2022learn,van2022thermodynamic,blom2024milestoning,yang2021physical}.
On the one hand, it suggests that continuous downscaling makes it impossible to optimize both the reliability and energy efficiency of ring clocks in the current technological paradigm \cite{theis2017end}. On the other hand, the fact that chemical clocks are loosely constrained by \eqref{bound_micro}, implies that $\mathcal{N}$ cannot be taken as a reliable proxy for the free energy dissipation of biological oscillators \cite{chen2024energy}, and other thermodynamic observables should be monitored (e.g., by calorimetry \cite{rodenfels2019heat,diterlizzi2024variance}).

\begin{acknowledgments}
    G.F. acknowledges funding from the European Union (NextGenerationEU) and the program STARS@UNIPD with project “ThermoComplex”.
\end{acknowledgments}

\bibliography{bibliography}

\begin{thebibliography}{71}
\expandafter\ifx\csname natexlab\endcsname\relax\def\natexlab#1{#1}\fi
\expandafter\ifx\csname bibnamefont\endcsname\relax
  \def\bibnamefont#1{#1}\fi
\expandafter\ifx\csname bibfnamefont\endcsname\relax
  \def\bibfnamefont#1{#1}\fi
\expandafter\ifx\csname citenamefont\endcsname\relax
  \def\citenamefont#1{#1}\fi
\expandafter\ifx\csname url\endcsname\relax
  \def\url#1{\texttt{#1}}\fi
\expandafter\ifx\csname urlprefix\endcsname\relax\def\urlprefix{URL }\fi
\providecommand{\bibinfo}[2]{#2}
\providecommand{\eprint}[2][]{\url{#2}}

\bibitem[{\citenamefont{Esposito et~al.}(2010)\citenamefont{Esposito, Kawai, Lindenberg, and Van~den Broeck}}]{esposito2010maximum}
\bibinfo{author}{\bibfnamefont{M.}~\bibnamefont{Esposito}}, \bibinfo{author}{\bibfnamefont{R.}~\bibnamefont{Kawai}}, \bibinfo{author}{\bibfnamefont{K.}~\bibnamefont{Lindenberg}}, \bibnamefont{and} \bibinfo{author}{\bibfnamefont{C.}~\bibnamefont{Van~den Broeck}}, \bibinfo{journal}{Phys. Rev. Lett.} \textbf{\bibinfo{volume}{105}}, \bibinfo{pages}{150603} (\bibinfo{year}{2010}), \urlprefix\url{https://link.aps.org/doi/10.1103/PhysRevLett.105.150603}.

\bibitem[{\citenamefont{Barato and Seifert}(2015)}]{Barato2015}
\bibinfo{author}{\bibfnamefont{A.~C.} \bibnamefont{Barato}} \bibnamefont{and} \bibinfo{author}{\bibfnamefont{U.}~\bibnamefont{Seifert}}, \bibinfo{journal}{Phys. Rev. Lett.} \textbf{\bibinfo{volume}{114}}, \bibinfo{pages}{158101} (\bibinfo{year}{2015}), \urlprefix\url{https://link.aps.org/doi/10.1103/PhysRevLett.114.158101}.

\bibitem[{\citenamefont{Gingrich et~al.}(2016)\citenamefont{Gingrich, Horowitz, Perunov, and England}}]{gingrich2016dissipation}
\bibinfo{author}{\bibfnamefont{T.~R.} \bibnamefont{Gingrich}}, \bibinfo{author}{\bibfnamefont{J.~M.} \bibnamefont{Horowitz}}, \bibinfo{author}{\bibfnamefont{N.}~\bibnamefont{Perunov}}, \bibnamefont{and} \bibinfo{author}{\bibfnamefont{J.~L.} \bibnamefont{England}}, \bibinfo{journal}{Physical review letters} \textbf{\bibinfo{volume}{116}}, \bibinfo{pages}{120601} (\bibinfo{year}{2016}).

\bibitem[{\citenamefont{Hasegawa and Van~Vu}(2019)}]{hasegawa2019fluctuation}
\bibinfo{author}{\bibfnamefont{Y.}~\bibnamefont{Hasegawa}} \bibnamefont{and} \bibinfo{author}{\bibfnamefont{T.}~\bibnamefont{Van~Vu}}, \bibinfo{journal}{Physical Review Letters} \textbf{\bibinfo{volume}{123}}, \bibinfo{pages}{110602} (\bibinfo{year}{2019}).

\bibitem[{\citenamefont{Falasco et~al.}(2020)\citenamefont{Falasco, Esposito, and Delvenne}}]{falasco2019unifying}
\bibinfo{author}{\bibfnamefont{G.}~\bibnamefont{Falasco}}, \bibinfo{author}{\bibfnamefont{M.}~\bibnamefont{Esposito}}, \bibnamefont{and} \bibinfo{author}{\bibfnamefont{J.-C.} \bibnamefont{Delvenne}}, \bibinfo{journal}{New J. Phys.} \textbf{\bibinfo{volume}{22}}, \bibinfo{pages}{053046} (\bibinfo{year}{2020}), ISSN \bibinfo{issn}{1367-2630}.

\bibitem[{\citenamefont{Horowitz and Gingrich}(2020)}]{horowitz2020thermodynamic}
\bibinfo{author}{\bibfnamefont{J.~M.} \bibnamefont{Horowitz}} \bibnamefont{and} \bibinfo{author}{\bibfnamefont{T.~R.} \bibnamefont{Gingrich}}, \bibinfo{journal}{Nat. Phys.} \textbf{\bibinfo{volume}{16}}, \bibinfo{pages}{15} (\bibinfo{year}{2020}).

\bibitem[{\citenamefont{Dechant and Sasa}(2020)}]{dechant2020}
\bibinfo{author}{\bibfnamefont{A.}~\bibnamefont{Dechant}} \bibnamefont{and} \bibinfo{author}{\bibfnamefont{S.-i.} \bibnamefont{Sasa}}, \bibinfo{journal}{Proc. Natl. Acad. Sci. U.S.A.} \textbf{\bibinfo{volume}{117}}, \bibinfo{pages}{6430} (\bibinfo{year}{2020}).

\bibitem[{\citenamefont{Gingrich and Horowitz}(2017)}]{gingrich2017fundamental}
\bibinfo{author}{\bibfnamefont{T.~R.} \bibnamefont{Gingrich}} \bibnamefont{and} \bibinfo{author}{\bibfnamefont{J.~M.} \bibnamefont{Horowitz}}, \bibinfo{journal}{Physical review letters} \textbf{\bibinfo{volume}{119}}, \bibinfo{pages}{170601} (\bibinfo{year}{2017}).

\bibitem[{\citenamefont{Falasco and Esposito}(2020)}]{falasco2020dissipation}
\bibinfo{author}{\bibfnamefont{G.}~\bibnamefont{Falasco}} \bibnamefont{and} \bibinfo{author}{\bibfnamefont{M.}~\bibnamefont{Esposito}}, \bibinfo{journal}{Physical Review Letters} \textbf{\bibinfo{volume}{125}}, \bibinfo{pages}{120604} (\bibinfo{year}{2020}).

\bibitem[{\citenamefont{Ptaszy{\'n}ski et~al.}(2024)\citenamefont{Ptaszy{\'n}ski, Aslyamov, and Esposito}}]{ptaszynski2024dissipation}
\bibinfo{author}{\bibfnamefont{K.}~\bibnamefont{Ptaszy{\'n}ski}}, \bibinfo{author}{\bibfnamefont{T.}~\bibnamefont{Aslyamov}}, \bibnamefont{and} \bibinfo{author}{\bibfnamefont{M.}~\bibnamefont{Esposito}}, \bibinfo{journal}{Physical Review Letters} \textbf{\bibinfo{volume}{133}}, \bibinfo{pages}{227101} (\bibinfo{year}{2024}).

\bibitem[{\citenamefont{Kwon et~al.}(2024)\citenamefont{Kwon, Chun, Park, and Lee}}]{kwon2024fluctuation}
\bibinfo{author}{\bibfnamefont{E.}~\bibnamefont{Kwon}}, \bibinfo{author}{\bibfnamefont{H.-M.} \bibnamefont{Chun}}, \bibinfo{author}{\bibfnamefont{H.}~\bibnamefont{Park}}, \bibnamefont{and} \bibinfo{author}{\bibfnamefont{J.~S.} \bibnamefont{Lee}}, \bibinfo{journal}{arXiv preprint arXiv:2411.18108}  (\bibinfo{year}{2024}).

\bibitem[{\citenamefont{Ptaszynski et~al.}(2024)\citenamefont{Ptaszynski, Aslyamov, and Esposito}}]{ptaszynski2024nonequilibrium}
\bibinfo{author}{\bibfnamefont{K.}~\bibnamefont{Ptaszynski}}, \bibinfo{author}{\bibfnamefont{T.}~\bibnamefont{Aslyamov}}, \bibnamefont{and} \bibinfo{author}{\bibfnamefont{M.}~\bibnamefont{Esposito}}, \bibinfo{journal}{arXiv preprint arXiv:2412.10233}  (\bibinfo{year}{2024}).

\bibitem[{\citenamefont{Shiraishi et~al.}(2018)\citenamefont{Shiraishi, Funo, and Saito}}]{shiraishi2018speed}
\bibinfo{author}{\bibfnamefont{N.}~\bibnamefont{Shiraishi}}, \bibinfo{author}{\bibfnamefont{K.}~\bibnamefont{Funo}}, \bibnamefont{and} \bibinfo{author}{\bibfnamefont{K.}~\bibnamefont{Saito}}, \bibinfo{journal}{Physical review letters} \textbf{\bibinfo{volume}{121}}, \bibinfo{pages}{070601} (\bibinfo{year}{2018}).

\bibitem[{\citenamefont{Vo et~al.}(2020)\citenamefont{Vo, Van~Vu, and Hasegawa}}]{vo2020unified}
\bibinfo{author}{\bibfnamefont{V.~T.} \bibnamefont{Vo}}, \bibinfo{author}{\bibfnamefont{T.}~\bibnamefont{Van~Vu}}, \bibnamefont{and} \bibinfo{author}{\bibfnamefont{Y.}~\bibnamefont{Hasegawa}}, \bibinfo{journal}{Physical Review E} \textbf{\bibinfo{volume}{102}}, \bibinfo{pages}{062132} (\bibinfo{year}{2020}).

\bibitem[{\citenamefont{Van~Vu and Saito}(2023)}]{van2023thermodynamic}
\bibinfo{author}{\bibfnamefont{T.}~\bibnamefont{Van~Vu}} \bibnamefont{and} \bibinfo{author}{\bibfnamefont{K.}~\bibnamefont{Saito}}, \bibinfo{journal}{Physical Review X} \textbf{\bibinfo{volume}{13}}, \bibinfo{pages}{011013} (\bibinfo{year}{2023}).

\bibitem[{\citenamefont{Dieball and Godec}(2024)}]{dieball2024thermodynamic}
\bibinfo{author}{\bibfnamefont{C.}~\bibnamefont{Dieball}} \bibnamefont{and} \bibinfo{author}{\bibfnamefont{A.}~\bibnamefont{Godec}}, \bibinfo{journal}{Physical Review Letters} \textbf{\bibinfo{volume}{133}}, \bibinfo{pages}{067101} (\bibinfo{year}{2024}).

\bibitem[{\citenamefont{Marsland~III et~al.}(2019)\citenamefont{Marsland~III, Cui, and Horowitz}}]{marsland2019}
\bibinfo{author}{\bibfnamefont{R.}~\bibnamefont{Marsland~III}}, \bibinfo{author}{\bibfnamefont{W.}~\bibnamefont{Cui}}, \bibnamefont{and} \bibinfo{author}{\bibfnamefont{J.~M.} \bibnamefont{Horowitz}}, \bibinfo{journal}{J. R. Soc. Interface} \textbf{\bibinfo{volume}{16}}, \bibinfo{pages}{20190098} (\bibinfo{year}{2019}).

\bibitem[{\citenamefont{Falasco et~al.}(2019)\citenamefont{Falasco, Cossetto, Penocchio, and Esposito}}]{Falasco2019a}
\bibinfo{author}{\bibfnamefont{G.}~\bibnamefont{Falasco}}, \bibinfo{author}{\bibfnamefont{T.}~\bibnamefont{Cossetto}}, \bibinfo{author}{\bibfnamefont{E.}~\bibnamefont{Penocchio}}, \bibnamefont{and} \bibinfo{author}{\bibfnamefont{M.}~\bibnamefont{Esposito}}, \bibinfo{journal}{New J. Phys.} \textbf{\bibinfo{volume}{21}}, \bibinfo{pages}{073005} (\bibinfo{year}{2019}), ISSN \bibinfo{issn}{1367-2630}.

\bibitem[{\citenamefont{Zheng and Tang}(2024)}]{zheng2024topological}
\bibinfo{author}{\bibfnamefont{C.}~\bibnamefont{Zheng}} \bibnamefont{and} \bibinfo{author}{\bibfnamefont{E.}~\bibnamefont{Tang}}, \bibinfo{journal}{Nature Communications} \textbf{\bibinfo{volume}{15}}, \bibinfo{pages}{6453} (\bibinfo{year}{2024}).

\bibitem[{\citenamefont{Cao et~al.}(2015)\citenamefont{Cao, Wang, Ouyang, and Tu}}]{cao2015free}
\bibinfo{author}{\bibfnamefont{Y.}~\bibnamefont{Cao}}, \bibinfo{author}{\bibfnamefont{H.}~\bibnamefont{Wang}}, \bibinfo{author}{\bibfnamefont{Q.}~\bibnamefont{Ouyang}}, \bibnamefont{and} \bibinfo{author}{\bibfnamefont{Y.}~\bibnamefont{Tu}}, \bibinfo{journal}{Nat. Phys.} \textbf{\bibinfo{volume}{11}}, \bibinfo{pages}{772} (\bibinfo{year}{2015}).

\bibitem[{\citenamefont{Oberreiter et~al.}(2022)\citenamefont{Oberreiter, Seifert, and Barato}}]{oberreiter2022universal}
\bibinfo{author}{\bibfnamefont{L.}~\bibnamefont{Oberreiter}}, \bibinfo{author}{\bibfnamefont{U.}~\bibnamefont{Seifert}}, \bibnamefont{and} \bibinfo{author}{\bibfnamefont{A.~C.} \bibnamefont{Barato}}, \bibinfo{journal}{Physical Review E} \textbf{\bibinfo{volume}{106}}, \bibinfo{pages}{014106} (\bibinfo{year}{2022}).

\bibitem[{\citenamefont{Ohga et~al.}(2023)\citenamefont{Ohga, Ito, and Kolchinsky}}]{ohga2023thermodynamic}
\bibinfo{author}{\bibfnamefont{N.}~\bibnamefont{Ohga}}, \bibinfo{author}{\bibfnamefont{S.}~\bibnamefont{Ito}}, \bibnamefont{and} \bibinfo{author}{\bibfnamefont{A.}~\bibnamefont{Kolchinsky}}, \bibinfo{journal}{Physical Review Letters} \textbf{\bibinfo{volume}{131}}, \bibinfo{pages}{077101} (\bibinfo{year}{2023}).

\bibitem[{\citenamefont{Shiraishi}(2023)}]{shiraishi2023entropy}
\bibinfo{author}{\bibfnamefont{N.}~\bibnamefont{Shiraishi}}, \bibinfo{journal}{Physical Review E} \textbf{\bibinfo{volume}{108}}, \bibinfo{pages}{L042103} (\bibinfo{year}{2023}).

\bibitem[{\citenamefont{Monti et~al.}(2018)\citenamefont{Monti, Lubensky, and ten Wolde}}]{Monti2018robustness}
\bibinfo{author}{\bibfnamefont{M.}~\bibnamefont{Monti}}, \bibinfo{author}{\bibfnamefont{D.~K.} \bibnamefont{Lubensky}}, \bibnamefont{and} \bibinfo{author}{\bibfnamefont{P.~R.} \bibnamefont{ten Wolde}}, \bibinfo{journal}{Phys. Rev. Lett.} \textbf{\bibinfo{volume}{121}}, \bibinfo{pages}{078101} (\bibinfo{year}{2018}), \urlprefix\url{https://link.aps.org/doi/10.1103/PhysRevLett.121.078101}.

\bibitem[{\citenamefont{Voorsluijs et~al.}(2024)\citenamefont{Voorsluijs, Avanzini, Falasco, Esposito, and Skupin}}]{voorsluijs2024calcium}
\bibinfo{author}{\bibfnamefont{V.}~\bibnamefont{Voorsluijs}}, \bibinfo{author}{\bibfnamefont{F.}~\bibnamefont{Avanzini}}, \bibinfo{author}{\bibfnamefont{G.}~\bibnamefont{Falasco}}, \bibinfo{author}{\bibfnamefont{M.}~\bibnamefont{Esposito}}, \bibnamefont{and} \bibinfo{author}{\bibfnamefont{A.}~\bibnamefont{Skupin}}, \bibinfo{journal}{Iscience} \textbf{\bibinfo{volume}{27}} (\bibinfo{year}{2024}).

\bibitem[{\citenamefont{Da~Dalt and Sheikholeslami}(2018)}]{dadalt2018understanding}
\bibinfo{author}{\bibfnamefont{N.}~\bibnamefont{Da~Dalt}} \bibnamefont{and} \bibinfo{author}{\bibfnamefont{A.}~\bibnamefont{Sheikholeslami}}, \emph{\bibinfo{title}{Understanding jitter and phase noise: A circuits and systems perspective}} (\bibinfo{publisher}{Cambridge University Press}, \bibinfo{year}{2018}).

\bibitem[{\citenamefont{Falasco and Esposito}(2025)}]{RevModPhys.97.015002}
\bibinfo{author}{\bibfnamefont{G.}~\bibnamefont{Falasco}} \bibnamefont{and} \bibinfo{author}{\bibfnamefont{M.}~\bibnamefont{Esposito}}, \bibinfo{journal}{Rev. Mod. Phys.} \textbf{\bibinfo{volume}{97}}, \bibinfo{pages}{015002} (\bibinfo{year}{2025}), \urlprefix\url{https://link.aps.org/doi/10.1103/RevModPhys.97.015002}.

\bibitem[{\citenamefont{Freitas et~al.}(2020)\citenamefont{Freitas, Delvenne, and Esposito}}]{Freitas2020linear}
\bibinfo{author}{\bibfnamefont{N.}~\bibnamefont{Freitas}}, \bibinfo{author}{\bibfnamefont{J.-C.} \bibnamefont{Delvenne}}, \bibnamefont{and} \bibinfo{author}{\bibfnamefont{M.}~\bibnamefont{Esposito}}, \bibinfo{journal}{Phys. Rev. X} \textbf{\bibinfo{volume}{10}}, \bibinfo{pages}{031005} (\bibinfo{year}{2020}), ISSN \bibinfo{issn}{2160-3308}.

\bibitem[{\citenamefont{Freitas et~al.}(2021{\natexlab{a}})\citenamefont{Freitas, Delvenne, and Esposito}}]{Freitas2020nonlinear}
\bibinfo{author}{\bibfnamefont{N.}~\bibnamefont{Freitas}}, \bibinfo{author}{\bibfnamefont{J.-C.} \bibnamefont{Delvenne}}, \bibnamefont{and} \bibinfo{author}{\bibfnamefont{M.}~\bibnamefont{Esposito}}, \bibinfo{journal}{Phys. Rev. X} \textbf{\bibinfo{volume}{11}}, \bibinfo{pages}{031064} (\bibinfo{year}{2021}{\natexlab{a}}), \urlprefix\url{https://link.aps.org/doi/10.1103/PhysRevX.11.031064}.

\bibitem[{\citenamefont{Gopal et~al.}(2022)\citenamefont{Gopal, Esposito, and Freitas}}]{gopal2022large}
\bibinfo{author}{\bibfnamefont{A.}~\bibnamefont{Gopal}}, \bibinfo{author}{\bibfnamefont{M.}~\bibnamefont{Esposito}}, \bibnamefont{and} \bibinfo{author}{\bibfnamefont{N.}~\bibnamefont{Freitas}}, \bibinfo{journal}{Physical Review B} \textbf{\bibinfo{volume}{106}}, \bibinfo{pages}{155303} (\bibinfo{year}{2022}).

\bibitem[{\citenamefont{Gopal et~al.}(2024)\citenamefont{Gopal, Esposito, and Freitas}}]{gopal2024thermodynamic}
\bibinfo{author}{\bibfnamefont{A.}~\bibnamefont{Gopal}}, \bibinfo{author}{\bibfnamefont{M.}~\bibnamefont{Esposito}}, \bibnamefont{and} \bibinfo{author}{\bibfnamefont{N.}~\bibnamefont{Freitas}}, \bibinfo{journal}{Physical Review B} \textbf{\bibinfo{volume}{109}}, \bibinfo{pages}{085421} (\bibinfo{year}{2024}).

\bibitem[{\citenamefont{Van~Kampen}(2007)}]{vanKampen}
\bibinfo{author}{\bibfnamefont{N.}~\bibnamefont{Van~Kampen}}, \emph{\bibinfo{title}{Stochastic Processes in Physics and Chemistry}} (\bibinfo{publisher}{North Holland}, \bibinfo{year}{2007}), ISBN \bibinfo{isbn}{9780444529657}.

\bibitem[{\citenamefont{Gillespie}(2001)}]{gillespie2001approximate}
\bibinfo{author}{\bibfnamefont{D.~T.} \bibnamefont{Gillespie}}, \bibinfo{journal}{The Journal of chemical physics} \textbf{\bibinfo{volume}{115}}, \bibinfo{pages}{1716} (\bibinfo{year}{2001}).

\bibitem[{\citenamefont{Lazarescu et~al.}(2019)\citenamefont{Lazarescu, Cossetto, Falasco, and Esposito}}]{lazarescu2019large}
\bibinfo{author}{\bibfnamefont{A.}~\bibnamefont{Lazarescu}}, \bibinfo{author}{\bibfnamefont{T.}~\bibnamefont{Cossetto}}, \bibinfo{author}{\bibfnamefont{G.}~\bibnamefont{Falasco}}, \bibnamefont{and} \bibinfo{author}{\bibfnamefont{M.}~\bibnamefont{Esposito}}, \bibinfo{journal}{The Journal of Chemical Physics} \textbf{\bibinfo{volume}{151}} (\bibinfo{year}{2019}).

\bibitem[{\citenamefont{Boland et~al.}(2008)\citenamefont{Boland, Galla, and McKane}}]{boland2008limit}
\bibinfo{author}{\bibfnamefont{R.~P.} \bibnamefont{Boland}}, \bibinfo{author}{\bibfnamefont{T.}~\bibnamefont{Galla}}, \bibnamefont{and} \bibinfo{author}{\bibfnamefont{A.~J.} \bibnamefont{McKane}}, \bibinfo{journal}{Journal of Statistical Mechanics: Theory and Experiment} \textbf{\bibinfo{volume}{2008}}, \bibinfo{pages}{P09001} (\bibinfo{year}{2008}).

\bibitem[{\citenamefont{Graham}(1987)}]{graham1987macroscopic}
\bibinfo{author}{\bibfnamefont{R.}~\bibnamefont{Graham}}, in \emph{\bibinfo{booktitle}{Fluctuations and Stochastic Phenomena in Condensed Matter}} (\bibinfo{publisher}{Springer}, \bibinfo{year}{1987}), pp. \bibinfo{pages}{1--34}.

\bibitem[{\citenamefont{Vance and Ross}(1996)}]{vance1996fluctuations}
\bibinfo{author}{\bibfnamefont{W.}~\bibnamefont{Vance}} \bibnamefont{and} \bibinfo{author}{\bibfnamefont{J.}~\bibnamefont{Ross}}, \bibinfo{journal}{The Journal of chemical physics} \textbf{\bibinfo{volume}{105}}, \bibinfo{pages}{479} (\bibinfo{year}{1996}).

\bibitem[{\citenamefont{Gaspard}(2002{\natexlab{a}})}]{gaspard2002clocks}
\bibinfo{author}{\bibfnamefont{P.}~\bibnamefont{Gaspard}}, \bibinfo{journal}{J. Chem. Phys.} \textbf{\bibinfo{volume}{117}}, \bibinfo{pages}{8905} (\bibinfo{year}{2002}{\natexlab{a}}).

\bibitem[{\citenamefont{Dykman et~al.}(1994)\citenamefont{Dykman, Mori, Ross, and Hunt}}]{dykman1994large}
\bibinfo{author}{\bibfnamefont{M.~I.} \bibnamefont{Dykman}}, \bibinfo{author}{\bibfnamefont{E.}~\bibnamefont{Mori}}, \bibinfo{author}{\bibfnamefont{J.}~\bibnamefont{Ross}}, \bibnamefont{and} \bibinfo{author}{\bibfnamefont{P.~M.} \bibnamefont{Hunt}}, \bibinfo{journal}{J. Chem. Phys.} \textbf{\bibinfo{volume}{100}}, \bibinfo{pages}{5735} (\bibinfo{year}{1994}), ISSN \bibinfo{issn}{0021-9606}.

\bibitem[{\citenamefont{Gaspard}(2002{\natexlab{b}})}]{gaspard2002trace}
\bibinfo{author}{\bibfnamefont{P.}~\bibnamefont{Gaspard}}, \bibinfo{journal}{Journal of statistical physics} \textbf{\bibinfo{volume}{106}}, \bibinfo{pages}{57} (\bibinfo{year}{2002}{\natexlab{b}}).

\bibitem[{\citenamefont{Remlein et~al.}(2022)\citenamefont{Remlein, Weissmann, and Seifert}}]{Remlein2022coherence}
\bibinfo{author}{\bibfnamefont{B.}~\bibnamefont{Remlein}}, \bibinfo{author}{\bibfnamefont{V.}~\bibnamefont{Weissmann}}, \bibnamefont{and} \bibinfo{author}{\bibfnamefont{U.}~\bibnamefont{Seifert}}, \bibinfo{journal}{Phys. Rev. E} \textbf{\bibinfo{volume}{105}}, \bibinfo{pages}{064101} (\bibinfo{year}{2022}), \urlprefix\url{https://link.aps.org/doi/10.1103/PhysRevE.105.064101}.

\bibitem[{\citenamefont{Remlein}(2024)}]{remlein2024nonlinear}
\bibinfo{author}{\bibfnamefont{B.}~\bibnamefont{Remlein}}, \bibinfo{type}{Doctoral dissertation}, \bibinfo{school}{University of Stuttgart} (\bibinfo{year}{2024}).

\bibitem[{\citenamefont{Sanders et~al.}(2007)\citenamefont{Sanders, Verhulst, and Murdock}}]{sanders2007averaging}
\bibinfo{author}{\bibfnamefont{J.~A.} \bibnamefont{Sanders}}, \bibinfo{author}{\bibfnamefont{F.}~\bibnamefont{Verhulst}}, \bibnamefont{and} \bibinfo{author}{\bibfnamefont{J.}~\bibnamefont{Murdock}}, \emph{\bibinfo{title}{Averaging methods in nonlinear dynamical systems}}, vol.~\bibinfo{volume}{59} (\bibinfo{publisher}{Springer}, \bibinfo{year}{2007}).

\bibitem[{\citenamefont{Hajimiri et~al.}(1999)\citenamefont{Hajimiri, Limotyrakis, and Lee}}]{Hajimiri1999}
\bibinfo{author}{\bibfnamefont{A.}~\bibnamefont{Hajimiri}}, \bibinfo{author}{\bibfnamefont{S.}~\bibnamefont{Limotyrakis}}, \bibnamefont{and} \bibinfo{author}{\bibfnamefont{T.}~\bibnamefont{Lee}}, \bibinfo{journal}{IEEE Journal of Solid-State Circuits} \textbf{\bibinfo{volume}{34}}, \bibinfo{pages}{790} (\bibinfo{year}{1999}).

\bibitem[{\citenamefont{Gopal}(2024)}]{gopal2024when}
\bibinfo{author}{\bibfnamefont{A.}~\bibnamefont{Gopal}}, \bibinfo{type}{Doctoral dissertation}, \bibinfo{school}{University of Luxembourg} (\bibinfo{year}{2024}).

\bibitem[{\citenamefont{Freitas et~al.}(2021{\natexlab{b}})\citenamefont{Freitas, Falasco, and Esposito}}]{Freitas2021linear}
\bibinfo{author}{\bibfnamefont{N.}~\bibnamefont{Freitas}}, \bibinfo{author}{\bibfnamefont{G.}~\bibnamefont{Falasco}}, \bibnamefont{and} \bibinfo{author}{\bibfnamefont{M.}~\bibnamefont{Esposito}}, \bibinfo{journal}{New J. Phys.} \textbf{\bibinfo{volume}{23}}, \bibinfo{pages}{093003} (\bibinfo{year}{2021}{\natexlab{b}}), ISSN \bibinfo{issn}{1367-2630}.

\bibitem[{\citenamefont{Blokhuis et~al.}(2020)\citenamefont{Blokhuis, Lacoste, and Nghe}}]{blokhuis2020universal}
\bibinfo{author}{\bibfnamefont{A.}~\bibnamefont{Blokhuis}}, \bibinfo{author}{\bibfnamefont{D.}~\bibnamefont{Lacoste}}, \bibnamefont{and} \bibinfo{author}{\bibfnamefont{P.}~\bibnamefont{Nghe}}, \bibinfo{journal}{Proceedings of the National Academy of Sciences} \textbf{\bibinfo{volume}{117}}, \bibinfo{pages}{25230} (\bibinfo{year}{2020}).

\bibitem[{\citenamefont{Wilhelm}(2009)}]{wilhelm2009smallest}
\bibinfo{author}{\bibfnamefont{T.}~\bibnamefont{Wilhelm}}, \bibinfo{journal}{BMC systems biology} \textbf{\bibinfo{volume}{3}}, \bibinfo{pages}{1} (\bibinfo{year}{2009}).

\bibitem[{\citenamefont{Prigogine and Lefever}(1968)}]{Prigogine1968}
\bibinfo{author}{\bibfnamefont{I.}~\bibnamefont{Prigogine}} \bibnamefont{and} \bibinfo{author}{\bibfnamefont{R.}~\bibnamefont{Lefever}}, \bibinfo{journal}{The Journal of Chemical Physics} \textbf{\bibinfo{volume}{48}}, \bibinfo{pages}{1695} (\bibinfo{year}{1968}), ISSN \bibinfo{issn}{0021-9606}, \eprint{https://pubs.aip.org/aip/jcp/article-pdf/48/4/1695/18854064/1695\_1\_online.pdf}, \urlprefix\url{https://doi.org/10.1063/1.1668896}.

\bibitem[{\citenamefont{Nguyen et~al.}(2018)\citenamefont{Nguyen, Seifert, and Barato}}]{nguyen2018phase}
\bibinfo{author}{\bibfnamefont{B.}~\bibnamefont{Nguyen}}, \bibinfo{author}{\bibfnamefont{U.}~\bibnamefont{Seifert}}, \bibnamefont{and} \bibinfo{author}{\bibfnamefont{A.~C.} \bibnamefont{Barato}}, \bibinfo{journal}{The Journal of chemical physics} \textbf{\bibinfo{volume}{149}} (\bibinfo{year}{2018}).

\bibitem[{\citenamefont{Falasco et~al.}(2018)\citenamefont{Falasco, Rao, and Esposito}}]{Falasco2018Turing}
\bibinfo{author}{\bibfnamefont{G.}~\bibnamefont{Falasco}}, \bibinfo{author}{\bibfnamefont{R.}~\bibnamefont{Rao}}, \bibnamefont{and} \bibinfo{author}{\bibfnamefont{M.}~\bibnamefont{Esposito}}, \bibinfo{journal}{Phys. Rev. Lett.} \textbf{\bibinfo{volume}{121}}, \bibinfo{pages}{108301} (\bibinfo{year}{2018}), \urlprefix\url{https://link.aps.org/doi/10.1103/PhysRevLett.121.108301}.

\bibitem[{\citenamefont{Avanzini et~al.}(2019)\citenamefont{Avanzini, Falasco, and Esposito}}]{Avanzini2019waves}
\bibinfo{author}{\bibfnamefont{F.}~\bibnamefont{Avanzini}}, \bibinfo{author}{\bibfnamefont{G.}~\bibnamefont{Falasco}}, \bibnamefont{and} \bibinfo{author}{\bibfnamefont{M.}~\bibnamefont{Esposito}}, \bibinfo{journal}{J. Chem. Phys.} \textbf{\bibinfo{volume}{151}}, \bibinfo{pages}{234103} (\bibinfo{year}{2019}), \urlprefix\url{https://doi.org/10.1063/1.5126528}.

\bibitem[{\citenamefont{Kish}(2002)}]{kish2002end}
\bibinfo{author}{\bibfnamefont{L.~B.} \bibnamefont{Kish}}, \bibinfo{journal}{Physics Letters A} \textbf{\bibinfo{volume}{305}}, \bibinfo{pages}{144} (\bibinfo{year}{2002}).

\bibitem[{\citenamefont{Gao and Limmer}(2021)}]{gao2021principles}
\bibinfo{author}{\bibfnamefont{C.~Y.} \bibnamefont{Gao}} \bibnamefont{and} \bibinfo{author}{\bibfnamefont{D.~T.} \bibnamefont{Limmer}}, \bibinfo{journal}{Physical Review Research} \textbf{\bibinfo{volume}{3}}, \bibinfo{pages}{033169} (\bibinfo{year}{2021}).

\bibitem[{\citenamefont{Freitas et~al.}(2022)\citenamefont{Freitas, Proesmans, and Esposito}}]{Freitas2021reliability}
\bibinfo{author}{\bibfnamefont{N.}~\bibnamefont{Freitas}}, \bibinfo{author}{\bibfnamefont{K.}~\bibnamefont{Proesmans}}, \bibnamefont{and} \bibinfo{author}{\bibfnamefont{M.}~\bibnamefont{Esposito}}, \bibinfo{journal}{Phys. Rev. E} \textbf{\bibinfo{volume}{105}}, \bibinfo{pages}{034107} (\bibinfo{year}{2022}), \urlprefix\url{https://link.aps.org/doi/10.1103/PhysRevE.105.034107}.

\bibitem[{\citenamefont{Konopik et~al.}(2023)\citenamefont{Konopik, Korten, Lutz, and Linke}}]{konopik2023fundamental}
\bibinfo{author}{\bibfnamefont{M.}~\bibnamefont{Konopik}}, \bibinfo{author}{\bibfnamefont{T.}~\bibnamefont{Korten}}, \bibinfo{author}{\bibfnamefont{E.}~\bibnamefont{Lutz}}, \bibnamefont{and} \bibinfo{author}{\bibfnamefont{H.}~\bibnamefont{Linke}}, \bibinfo{journal}{Nature Communications} \textbf{\bibinfo{volume}{14}}, \bibinfo{pages}{447} (\bibinfo{year}{2023}).

\bibitem[{\citenamefont{Wolpert et~al.}(2024)\citenamefont{Wolpert, Korbel, Lynn, Tasnim, Grochow, Karde{\c{s}}, Aimone, Balasubramanian, De~Giuli, Doty et~al.}}]{wolpert2024stochastic}
\bibinfo{author}{\bibfnamefont{D.~H.} \bibnamefont{Wolpert}}, \bibinfo{author}{\bibfnamefont{J.}~\bibnamefont{Korbel}}, \bibinfo{author}{\bibfnamefont{C.~W.} \bibnamefont{Lynn}}, \bibinfo{author}{\bibfnamefont{F.}~\bibnamefont{Tasnim}}, \bibinfo{author}{\bibfnamefont{J.~A.} \bibnamefont{Grochow}}, \bibinfo{author}{\bibfnamefont{G.}~\bibnamefont{Karde{\c{s}}}}, \bibinfo{author}{\bibfnamefont{J.~B.} \bibnamefont{Aimone}}, \bibinfo{author}{\bibfnamefont{V.}~\bibnamefont{Balasubramanian}}, \bibinfo{author}{\bibfnamefont{E.}~\bibnamefont{De~Giuli}}, \bibinfo{author}{\bibfnamefont{D.}~\bibnamefont{Doty}}, \bibnamefont{et~al.}, \bibinfo{journal}{Proceedings of the National Academy of Sciences} \textbf{\bibinfo{volume}{121}}, \bibinfo{pages}{e2321112121} (\bibinfo{year}{2024}).

\bibitem[{\citenamefont{Li et~al.}(2019)\citenamefont{Li, Horowitz, Gingrich, and Fakhri}}]{li2019quantifying}
\bibinfo{author}{\bibfnamefont{J.}~\bibnamefont{Li}}, \bibinfo{author}{\bibfnamefont{J.~M.} \bibnamefont{Horowitz}}, \bibinfo{author}{\bibfnamefont{T.~R.} \bibnamefont{Gingrich}}, \bibnamefont{and} \bibinfo{author}{\bibfnamefont{N.}~\bibnamefont{Fakhri}}, \bibinfo{journal}{Nature communications} \textbf{\bibinfo{volume}{10}}, \bibinfo{pages}{1666} (\bibinfo{year}{2019}).

\bibitem[{\citenamefont{Seifert}(2019)}]{seifert2019stochastic}
\bibinfo{author}{\bibfnamefont{U.}~\bibnamefont{Seifert}}, \bibinfo{journal}{Annual Review of Condensed Matter Physics} \textbf{\bibinfo{volume}{10}}, \bibinfo{pages}{171} (\bibinfo{year}{2019}).

\bibitem[{\citenamefont{Skinner and Dunkel}(2021)}]{skinner2021estimating}
\bibinfo{author}{\bibfnamefont{D.~J.} \bibnamefont{Skinner}} \bibnamefont{and} \bibinfo{author}{\bibfnamefont{J.}~\bibnamefont{Dunkel}}, \bibinfo{journal}{Physical review letters} \textbf{\bibinfo{volume}{127}}, \bibinfo{pages}{198101} (\bibinfo{year}{2021}).

\bibitem[{\citenamefont{Harunari et~al.}(2022)\citenamefont{Harunari, Dutta, Polettini, and Rold{\'a}n}}]{harunari2022learn}
\bibinfo{author}{\bibfnamefont{P.~E.} \bibnamefont{Harunari}}, \bibinfo{author}{\bibfnamefont{A.}~\bibnamefont{Dutta}}, \bibinfo{author}{\bibfnamefont{M.}~\bibnamefont{Polettini}}, \bibnamefont{and} \bibinfo{author}{\bibfnamefont{{\'E}.}~\bibnamefont{Rold{\'a}n}}, \bibinfo{journal}{Physical Review X} \textbf{\bibinfo{volume}{12}}, \bibinfo{pages}{041026} (\bibinfo{year}{2022}).

\bibitem[{\citenamefont{Van~der Meer et~al.}(2022)\citenamefont{Van~der Meer, Ertel, and Seifert}}]{van2022thermodynamic}
\bibinfo{author}{\bibfnamefont{J.}~\bibnamefont{Van~der Meer}}, \bibinfo{author}{\bibfnamefont{B.}~\bibnamefont{Ertel}}, \bibnamefont{and} \bibinfo{author}{\bibfnamefont{U.}~\bibnamefont{Seifert}}, \bibinfo{journal}{Physical Review X} \textbf{\bibinfo{volume}{12}}, \bibinfo{pages}{031025} (\bibinfo{year}{2022}).

\bibitem[{\citenamefont{Blom et~al.}(2024)\citenamefont{Blom, Song, Vouga, Godec, and Makarov}}]{blom2024milestoning}
\bibinfo{author}{\bibfnamefont{K.}~\bibnamefont{Blom}}, \bibinfo{author}{\bibfnamefont{K.}~\bibnamefont{Song}}, \bibinfo{author}{\bibfnamefont{E.}~\bibnamefont{Vouga}}, \bibinfo{author}{\bibfnamefont{A.}~\bibnamefont{Godec}}, \bibnamefont{and} \bibinfo{author}{\bibfnamefont{D.~E.} \bibnamefont{Makarov}}, \bibinfo{journal}{Proceedings of the National Academy of Sciences} \textbf{\bibinfo{volume}{121}}, \bibinfo{pages}{e2318333121} (\bibinfo{year}{2024}).

\bibitem[{\citenamefont{Yang et~al.}(2021)\citenamefont{Yang, Heinemann, Howard, Huber, Iyer-Biswas, Treut, Lynch, Montooth, Needleman, Pigolotti et~al.}}]{yang2021physical}
\bibinfo{author}{\bibfnamefont{X.}~\bibnamefont{Yang}}, \bibinfo{author}{\bibfnamefont{M.}~\bibnamefont{Heinemann}}, \bibinfo{author}{\bibfnamefont{J.}~\bibnamefont{Howard}}, \bibinfo{author}{\bibfnamefont{G.}~\bibnamefont{Huber}}, \bibinfo{author}{\bibfnamefont{S.}~\bibnamefont{Iyer-Biswas}}, \bibinfo{author}{\bibfnamefont{G.~L.} \bibnamefont{Treut}}, \bibinfo{author}{\bibfnamefont{M.}~\bibnamefont{Lynch}}, \bibinfo{author}{\bibfnamefont{K.~L.} \bibnamefont{Montooth}}, \bibinfo{author}{\bibfnamefont{D.~J.} \bibnamefont{Needleman}}, \bibinfo{author}{\bibfnamefont{S.}~\bibnamefont{Pigolotti}}, \bibnamefont{et~al.}, \bibinfo{journal}{Proceedings of the National Academy of Sciences} \textbf{\bibinfo{volume}{118}}, \bibinfo{pages}{e2026786118} (\bibinfo{year}{2021}), \urlprefix\url{https://www.pnas.org/doi/abs/10.1073/pnas.2026786118}.

\bibitem[{\citenamefont{Theis and Wong}(2017)}]{theis2017end}
\bibinfo{author}{\bibfnamefont{T.~N.} \bibnamefont{Theis}} \bibnamefont{and} \bibinfo{author}{\bibfnamefont{H.-S.~P.} \bibnamefont{Wong}}, \bibinfo{journal}{Computing in science \& engineering} \textbf{\bibinfo{volume}{19}}, \bibinfo{pages}{41} (\bibinfo{year}{2017}).

\bibitem[{\citenamefont{Chen et~al.}(2024)\citenamefont{Chen, Seara, Michaud, Kim, Bement, and Murrell}}]{chen2024energy}
\bibinfo{author}{\bibfnamefont{S.}~\bibnamefont{Chen}}, \bibinfo{author}{\bibfnamefont{D.~S.} \bibnamefont{Seara}}, \bibinfo{author}{\bibfnamefont{A.}~\bibnamefont{Michaud}}, \bibinfo{author}{\bibfnamefont{S.}~\bibnamefont{Kim}}, \bibinfo{author}{\bibfnamefont{W.~M.} \bibnamefont{Bement}}, \bibnamefont{and} \bibinfo{author}{\bibfnamefont{M.~P.} \bibnamefont{Murrell}}, \bibinfo{journal}{Nature Physics} pp. \bibinfo{pages}{1--9} (\bibinfo{year}{2024}).

\bibitem[{\citenamefont{Rodenfels et~al.}(2019)\citenamefont{Rodenfels, Neugebauer, and Howard}}]{rodenfels2019heat}
\bibinfo{author}{\bibfnamefont{J.}~\bibnamefont{Rodenfels}}, \bibinfo{author}{\bibfnamefont{K.~M.} \bibnamefont{Neugebauer}}, \bibnamefont{and} \bibinfo{author}{\bibfnamefont{J.}~\bibnamefont{Howard}}, \bibinfo{journal}{Developmental cell} \textbf{\bibinfo{volume}{48}}, \bibinfo{pages}{646} (\bibinfo{year}{2019}).

\bibitem[{\citenamefont{Di~Terlizzi et~al.}(2024)\citenamefont{Di~Terlizzi, Gironella, Herr{\'a}ez-Aguilar, Monroy, Baiesi, and Ritort}}]{diterlizzi2024variance}
\bibinfo{author}{\bibfnamefont{I.}~\bibnamefont{Di~Terlizzi}}, \bibinfo{author}{\bibfnamefont{M.}~\bibnamefont{Gironella}}, \bibinfo{author}{\bibfnamefont{D.}~\bibnamefont{Herr{\'a}ez-Aguilar}}, \bibinfo{author}{\bibfnamefont{F.}~\bibnamefont{Monroy}}, \bibinfo{author}{\bibfnamefont{M.}~\bibnamefont{Baiesi}}, \bibnamefont{and} \bibinfo{author}{\bibfnamefont{F.}~\bibnamefont{Ritort}}, \bibinfo{journal}{Science} \textbf{\bibinfo{volume}{383}}, \bibinfo{pages}{971} (\bibinfo{year}{2024}).

\bibitem[{\citenamefont{Xiao et~al.}(2007)\citenamefont{Xiao, Ma, Hou, and Xin}}]{xiao2007effects}
\bibinfo{author}{\bibfnamefont{T.}~\bibnamefont{Xiao}}, \bibinfo{author}{\bibfnamefont{J.}~\bibnamefont{Ma}}, \bibinfo{author}{\bibfnamefont{Z.}~\bibnamefont{Hou}}, \bibnamefont{and} \bibinfo{author}{\bibfnamefont{H.}~\bibnamefont{Xin}}, \bibinfo{journal}{New Journal of Physics} \textbf{\bibinfo{volume}{9}}, \bibinfo{pages}{403} (\bibinfo{year}{2007}).

\bibitem[{\citenamefont{Perko}(2013)}]{perko2013differential}
\bibinfo{author}{\bibfnamefont{L.}~\bibnamefont{Perko}}, \emph{\bibinfo{title}{Differential equations and dynamical systems}}, vol.~\bibinfo{volume}{7} (\bibinfo{publisher}{Springer Science \& Business Media}, \bibinfo{year}{2013}).

\bibitem[{\citenamefont{Andersen and Geer}(1982)}]{andersen1982power}
\bibinfo{author}{\bibfnamefont{C.}~\bibnamefont{Andersen}} \bibnamefont{and} \bibinfo{author}{\bibfnamefont{J.~F.} \bibnamefont{Geer}}, \bibinfo{journal}{SIAM Journal on Applied Mathematics} \textbf{\bibinfo{volume}{42}}, \bibinfo{pages}{678} (\bibinfo{year}{1982}).

\end{thebibliography}

\appendix
\onecolumngrid

\renewcommand{\thesection}{S\Roman{section}}
\renewcommand{\thefigure}{S\arabic{figure}}
\renewcommand{\thetable}{S\arabic{table}}
\renewcommand{\theequation}{S\arabic{equation}}
\setcounter{section}{0}
\setcounter{figure}{0}
\setcounter{table}{0}
\setcounter{equation}{0}

\newpage
\begin{center}
\textbf{\Large Supplementary Material}\\ 
\end{center}

\section{Dynamics tangent to the limit cycle}

     We provide here a detailed calculation of the projected dynamics along the limit cycle $x^*(t)$ along the lines of the derivation in \cite{Remlein2022coherence}. We make use of the unit vector tangent to the limit cycle 
        \begin{align}
             u(t):=\frac{F(\mathcal{x}^*(t))}{|F(\mathcal{x}^*(t))|},
        \end{align}
     which is well defined as the drift field is always non-vanishing on the cycle. Thanks to it, we can define the stochastic variable $x_{\parallel}(t):= x(t) \cdot u(t)$. We compute its time derivative by means of the Leibniz rule to obtain Eq. \mainref{Langevin_parallel} of the Letter:
        \begin{align}
            d_t x_{\parallel} &=  u\cdot d_t\,x + \left( d_t\,u\right)\cdot x  \\
            &= u\cdot F + \left( d_t\,u\right)\cdot x +u \cdot \sqrt{2\epsilon}\xi \\
            &= |F| + g(x_\parallel, x_\perp) +\sqrt{2D_\parallel\epsilon} \eta.
            \label{eq Leibniz}
        \end{align}

   Here we have defined
        \begin{align}
            g &:= \left( d_t\,u\right)\cdot x + u \cdot [F(x)-F(\mathcal{x}^*)] \\
            &=\left( d_t\,u\right)\cdot x + u_j [x_i-\mathcal{x}_i^*] \cdot \partial_i F_j(\mathcal{x}^*) +O(\epsilon),
        \end{align}
   where the equality follows from expanding the drift $F(x)$ around the limit cycle at the leading order in $|x-\mathcal{x}^*|  \sim \epsilon^{1/2}$. We have also defined the diffusion constant $D_\parallel:=  u_i u_j D_{ij}$ and the zero-mean Gaussian variable $\eta$ with correlations $\mean{\eta(t)\eta(0)}= \delta(t)$. Einstein's convention on repeated indices is assumed and we use the shorthand $\partial_i\equiv \partial/\partial x_i$.
   The diffusion constant $D_\parallel$ has been obtained requiring that $u \cdot \xi$ and $\sqrt{D_\parallel} \eta$ have the same statistics, i.e., that their correlation functions are equal:
        \begin{align}
            u_i(t)u_j(0) \mean{ \xi_i(t) \xi_j(0)} = u_i(t)u_j(0) D_{ij}(t) \delta(t)= u_i(t)u_j(t) D_{ij}(t) \delta(t) \underset{!}{=} D_\parallel(t) \mean{\eta(t)\eta(0)}.
        \end{align}
    
    In the following we work out the explicit for of $g(x_\parallel, x_\perp)$, considering the derivative
        \begin{equation}
            x \cdot \dfrac{d}{dt}u = x_i\,\dfrac{d}{dt}u^i = x_i\,\dot{\mathcal{x}}^*_j\,\partial_j\,u^i = x_i\,F_j\partial_j\dfrac{F^i}{|F|}.
            \label{eq dfPARdt}
        \end{equation}
    The derivative of the drift in \eqref{eq dfPARdt} reads
        \begin{align}
            \partial_j\dfrac{F^i}{|F|} =& \dfrac{(\partial_j\,F^i)|F|-F^i\partial_j|F|}{|F|^2} \\
            =&\, \dfrac{1}{|F|}\partial_jF^i - \dfrac{F^i}{|F|}\dfrac{\partial_j|F|}{|F|} \\
            =&\, \dfrac{1}{|F|}\left[J_{ij} - u^i\,\partial_j(F_kF^k)^{1/2}\right] 
        \end{align}
    having defined the Jacobian $J_{ij}(t):=\partial_jF_i(\mathcal{x}^*(t))$. Proceeding with the calculation,
        \begin{align}
            \partial_j\dfrac{F^i}{|F|}   =&\, \dfrac{1}{|F|}\left[J_{ij} - u^i\,\dfrac{1}{2}\,(F_kF^k)^{-1/2}\partial_j(F_kF^k)\right] \\
            =&\, \dfrac{1}{|F|}\left[J_{ij} - u^i\,(F_kF^k)^{-1/2}F_k\partial_jF^k\right] \\
            =&\, \dfrac{1}{|F|}\left[J_{ij} - u^i\,\dfrac{1}{|F|}F_k\partial_jF^k\right] \\
            =&\, \dfrac{1}{|F|}\left[  J_{ij} - u^i\,J_{jk}u^k \right] \\
            =&\, \dfrac{1}{|F|}  \underbrace{\left[\delta^{ik} - u^i\,u^k\right]}_{u_{\perp}^{ik}}  J_{jk},
        \end{align}
     we end up identifying the matrix $u_{\perp}^{ik}$, which is the projector onto the $d-1$ dimensional subspace orthogonal to the limit cycle. Plugging this expression into \eqref{eq dfPARdt}, and substituting $\dot{x}_j$ with its corresponding Langevin equation \mainref{Langevin} in the Letter, we obtain
        \begin{align}
            x_i F_j\,\partial_i\dfrac{F^i}{|F|} =&\, x_i\dfrac{F_j}{|F|}  u_{\perp}^{ik}  J_{jk} \\
            =&\, x_i  u_{\perp}^{ik}  u_j J_{jk}.
        \end{align}
    If we decompose the vector $x_i$ into its parallel and orthogonal components 
        \begin{equation}
            x_i = x_{\parallel}u_i + x_{\perp}^j u_{\perp}^{ji}
        \end{equation}
    and recalling that $u_i u_{\perp}^{ik}=0$, $x_{\perp}^j u_{\perp}^{jk}=x_{\perp}^k$ and $u_{\perp}^{ij}u_{\perp}^{jk} = u_{\perp}^{ik}$, we arrive at
        \begin{align}
            x\cdot \dfrac{d}{dt}u = x_{\perp}^k u^j J_{jk}.
            \label{eq g=0}
        \end{align}
    The function $g$ then reads 
        \begin{align}
            g= x_{\perp}^k u^j J_{jk} + [x_i(t)-\mathcal{x}_i^*(t)] u_j  J_{ij} +O(\epsilon),
        \end{align}
    and its average value reads at leading order in         $\epsilon \to 0$,
        \begin{align}
            \mean{g} \underset{\epsilon \to 0}{=} \mean{x_{\perp}^k} u^j J_{jk} .
        \end{align}
    We thus see that $g$ couples the tangent dynamics to transverse fluctuations. 
    In our proof we set $g=0$ to obtain an auxiliary, i.e., fictitious dynamics that we exploit to relate $\mathcal{N}^{(1)}$ to $\Sigma^{(1)}$.
    
    The computation of the diffusivity $D_{\parallel}$ appearing in \eqref{eq Leibniz} is left for the next section.

\section{Relation between number of coherent cycles and dissipation in the auxiliary dynamics}
    
    We justify expression \mainref{bound_N1} in the main text. Multiplying the integral in such expression by $t_p/t_p$ we can write down the expression for the time average over a period for the inverse of the entropy production rate in the auxiliary dynamics, namely
        \begin{equation}
            \dfrac{t_p}{t_p}\int_{0}^{t_p}dt\,\dfrac{1}{\dot{\Sigma}^{(1)}} = t_p \,\overline{\dfrac{1}{\dot{\Sigma}^{(1)}}}.
        \end{equation}
    Notice that along a limit cycle the entropy production will be strictly positive, so the inverse ratio of such a quantity is always well defined.

    The objective is to show that 
        \begin{equation}\label{eq AMvsHM}
            \overline{\dfrac{1}{\dot{\Sigma}^{(1)}}} \geq \dfrac{1}{\overline{\dot{\Sigma}^{(1)}}}
        \end{equation}
    which allows to make sense of $\mathcal{N}^{(1)}\leq \Sigma^{(1)}/2\pi$ reported in \mainref{bound_N1}. To do so, we imagine to discretize the time average as
        \begin{equation}
            \dfrac{1}{t_p}\int_0^{t_p} dt\,\dfrac{1}{\dot{\Sigma}(t)^{(1)}} \approx \dfrac{1}{N}\sum_{i=1}^N \dfrac{1}{\dot{\Sigma}^{(1)}_i} =AM\left(\{1/\dot{\Sigma}^{(1)}_i \}\right).
        \end{equation}
    with $\dot{\Sigma}^{(1)}(t)$ , $0 \leq t \leq t_p$ broken down into $\{\dot{\Sigma}^{(1)}_i\}_{i=1...N}$, with $N=t_p/\Delta t$ and $\Delta t \to 0$.
    
    The summation we wrote is the arithmetic mean ($AM$) of the reciprocals of $\dot{\Sigma}_i$, which we name with a short-hand notation. Such an average can be reformulated in terms of the harmonic mean ($HM$), given that the latter is defined as the reciprocal of the arithmetic mean of the reciprocal elements of the sample This fact leads to the next expression
        \begin{align}
            \dfrac{1}{N}\sum_{i=1}^N \dfrac{1}{\dot{\Sigma}^{(1)}_i} = AM\left(\bigg\{\dfrac{1}{\dot{\Sigma}^{(1)}_i}\bigg\}\right) = \dfrac{1}{HM\left(\{\dot{\Sigma}^{(1)}_i\}\right)} \geq \dfrac{1}{AM\left(\{\dot{\Sigma}^{(1)}_i\}\right)} = \dfrac{1}{\frac{1}{N}\sum_{i=1}^N \dot{\Sigma}^{(1)}_i}
            \label{eq inequality_of_means}
        \end{align}
    having exploited the hierarchy of Pythagorean means which tells that the harmonic mean is always less or equal than the arithmetic mean of the same set of elements. 
    
    At this point, we can resume the continuous time notation to write \eqref{eq inequality_of_means} as
        \begin{equation}
            \dfrac{1}{t_p}\int_0^{t_p} dt\,\dfrac{1}{\dot{\Sigma}(t)^{(1)}} \geq \left( 
            \dfrac{1}{t_p}\int_0^{t_p} dt\,\dot{\Sigma}(t)^{(1)}) \right)^{-1},
        \end{equation}
    which is what we need to motivate our conjectured inequality. 

\section{Relation between the dissipation functions of the and auxiliary and full dynamics}

    The purpose of this section is to provide an explicit proof for the inequality
        \begin{equation}
            \dot{\Sigma} = F_i D_{ij}^{-1}F_j\geq 
            |F|^2/D_{\parallel}
            = \dot{\Sigma}^{(1)}
        \end{equation}
   Our strategy exploits the Cauchy-Schwartz inequality
        \begin{equation}
            \dfrac{\left( \sum_i w_i v_i \right)^2}{\sum_i w_i^2} \leq \sum_i v_i^2 \quad, \quad w_i, v_i \in \mathbb{R}
            \label{eq CS}
        \end{equation}
    identifying $v_i, w_i$ with
        \begin{equation}
            w_i = \sum_{k} D_{ik}^{1/2}F_k \quad, \quad v_i = \sum_{j} F_j D_{ji}^{-1/2}
        \end{equation}
    in order to show our result. Thus, we compute the sum of the product, setting $G\equiv D^{1/2}$:
        \begin{align}
            \sum_{i}v_iw_i &= \sum_i \left(\sum_{j} F_j D_{ji}^{-1/2} \sum_{k} D_{ik}^{1/2}F_k \right) \\
            &= \sum_{jk}\left( \sum_i F_j D_{ji}^{-1/2} D_{ik}^{1/2}F_k \right) \\
            &= \sum_{jk}\left( \sum_i F_j G_{ji}^{-1} G_{ik}F_k \right) \, ,  \\
            &= \sum_{jk}\left( \sum_i F_j \delta_{jk}\,F_k \right)  \\
            &= \sum_j F_j\,F_j \\
            &= |F|^2
        \end{align}
    having used condition $G_{ji}^{-1}G_{ik} = \delta_{jk}$. The sum
        \begin{align}
            \sum_i v_i^2 &= \sum_i \left(\sum_{j} F_j D_{ji}^{-1/2} \sum_{k} F_k D_{ki}^{-1/2} \right) \\
            &= \sum_{j,k} \left(\sum_i F_j D_{ji}^{-1/2}D_{ki}^{-1/2}  F_k \right) \\
            &= \sum_{j,k}  F_j D_{jk}^{-1}  F_k = \epsilon \dot{\Sigma}
        \end{align}
    is the entropy production rate associated to the $d$-dimensional Langevin dynamics. In the last line we exploited $\sum_i C_{ji}C_{ik} = C_{jk}^2 = D_{jk}^{-1}$, with $C\equiv D^{-1/2}$. Analogously, we calculate the second sum
        \begin{equation}
            \sum_i w_i^2 = \sum_{j,k}  F_j D_{jk}  F_k .
        \end{equation}
    By plugging these results into the Cauchy-Schwarz inequality \eqref{eq CS}, recalling the definitions of $D_{\parallel}$ in terms of the tanget vector $u$,
        \begin{align}
            \dfrac{\left( \sum_i w_i v_i \right)^2}{\sum_i w_i^2} &= \dfrac{(|F|^2)^2}{ \sum_{j,k}  F_j D_{jk}  F_k} = \dfrac{|F|^2}{ \sum_{j,k}  \frac{F_j}{|F|} D_{jk}  \frac{F_k}{|F|}} \nonumber\\
            &= \dfrac{|F|^2}{ \sum_{j,k}  u_j D_{jk}  u_k} = \dfrac{|F|^2}{D_{\parallel}} = \epsilon \dot{\Sigma}^{(1)}.
        \end{align}
    The latter quantity is the entropy production rate associated to the auxiliary 1-dimensional dynamics. Therefore, \eqref{eq CS} becomes
        \begin{equation}
            \dot{\Sigma}^{(1)} \leq \dot{\Sigma}.
        \end{equation}
    With this computation we show how in the diffusive dynamics along the attractor we 'lose' part of the information about the dissipation, and this loss exists at every instant, i.e. the inequality holds the rates of entropy production.

\section{Saturation of the bound}

    We consider the stochastic normal form of the Hopf bifurcation (see, e.g., \cite{xiao2007effects}) 
        \begin{align}
            \dot r &=  \alpha r - \beta r^3 + \sqrt{2\epsilon} \xi_1 \label{eq HopfNormal1}\\
            \dot \theta &= \omega + \delta r^2  + \sqrt{2\epsilon/r^2}\xi_2 \label{eq HopfNormal2}
        \end{align}
    where $x_1 \equiv r$ is the radial coordinate, $x_2 \equiv \theta$ is the periodic angular coordinate and $\xi_i$ are zero-mean Gaussian noises with correlation $\mean{\xi_i(t) \xi_j(0)}= c \delta(t) \delta_{ij}$. For $\epsilon=0$ and $\alpha, \beta> 0$, this system displays the limit cycle solution $\mathcal{x}^{*}(t)=(r_*, t[\omega  + \delta r_*^2] )$ with radius $r_*=\sqrt{\alpha/\beta}$. With this in mind, we follow the strategy outlined in the Letter and project the dynamics along $\mathcal{x}^{*}(t)$ and evaluate the quantities entering \mainref{series_bounds}.

    The tangent vector $u$ is obtain from the drift field evaluated on the limit cycle reads $F(\mathcal{x}^{*}(t)) = ( 0 , \omega + \delta r_*^2 )$, whose norm reads $|F(\mathcal{x}^{*}(t))| = \omega + \delta r_*^2$, providing
        \begin{equation}
            u = \dfrac{F(\mathcal{x}^{*}(t))}{|F(\mathcal{x}^{*}(t))|} = 
           (0,1).
        \end{equation}
    We thus obtain the diffusivity $D_{\parallel} = D : uu=  d \beta/\alpha$, where the diffusion matrix for our system is given by $D = c\,\text{diag}(1, 1/r^2)$. Notice than  including nondiagonal elements in $D$ would not change this result. Indeed, given the form of the tangent vector $u$, only the element $D_{22}$ would replace $c$. 

    At this point, we can write down the Fokker-Planck equation associated to the tangent dynamics, describing the probability density function of the angle $\theta$ as $\epsilon \to 0$
        \begin{equation}
            \partial_t p(\theta, t) = -\partial _{\theta} \left(\omega + \frac{\delta\alpha}{\beta}\right)p(\theta,t) + \epsilon \dfrac{\beta c}{\alpha}\partial_{\theta}^2 p(\theta,t)
        \end{equation}
    which is solved by the Gaussian propagator
        \begin{align}
            p(\theta,t|\theta_0,0)=\frac{1}{\sqrt{4\pi t c \beta/(\Omega\alpha)}}\exp\left( -\frac{1}{2}\frac{(\theta-(\omega + \delta\alpha/\beta)t - \theta_0)}{2tc\beta/(\Omega\alpha)} \right).
        \end{align}
    This probability density allows us to evaluate the angle correlation, from which we directly extract the decorrelation time $\tau^{(1)}$ for the tangent dynamics. In fact, assuming a $\delta$ function as initial condition $p(\theta_0)=\delta(\theta_0-\theta(0))$, then the correlation function reads
        \begin{equation}
            \langle \theta(t)\theta_0 \rangle= \dfrac{\alpha}{\beta} \cos\theta_0 \cos((\omega + \delta\alpha/\beta)t + \theta_0)\mathrm{e}^{-\epsilon\frac{\beta c}{\alpha}t}
        \end{equation}
    which yields $\tau^{(1)}=D_{\parallel}^{-1}$ and thus $\mathcal{N}^{(1)}=\omega D_{\parallel}^{-1}$. On the other hand, the entropy production per cycle for the tangent dynamics can be calculated analytically as
        \begin{equation}
            \Sigma^{(1)} = \int_0^{t_p}dt\, \dfrac{|F(\mathcal{x}^{*}(t))|^2}{D_{\theta}} = t_p \dfrac{(\omega + \delta\alpha/\beta)^2}{\beta c}\alpha,
            \label{eq SigmaHopfNormal}
        \end{equation}
    with $t_p=2 \pi/\omega$.
    The peculiarity of this example is that for $\delta=0$, the angular dynamics, Eqs. \eqref{eq HopfNormal2}, is decoupled from \eqref{eq HopfNormal1}, and coincides with the auxiliary dynamics used in the Letter. As a result Eqs. \eqref{eq HopfNormal1}, \eqref{eq HopfNormal2} for $\delta=0$ saturate the bound \mainref {series_bounds}, because $\mathcal{N}^{(1)}=\mathcal{N}$ and $ \Sigma^{(1)}= \Sigma $.
    With this observation, we can show that \mainref{series_bounds} turns into a series of equalities starting from Eq. \eqref{eq SigmaHopfNormal}
        \begin{equation}
            \Sigma = \Sigma^{(1)} = \dfrac{\omega^2 \alpha}{\beta c} =2\pi\dfrac{\omega\alpha}{\beta c}  
            =2\pi \omega \tau =\mathcal{N}^{(1)} = \mathcal{N}
        \end{equation}

\section{Bistable overdamped system under rotational flow}

    \subsection{\emph{Mapping onto the Van der Pol oscillator and asymptotics}}\label{sec VdP}

        We consider the overdamped Langevin equation for the 2-dimensional coordinate $(x_1,x_2) \equiv( x,y)$ of a single particle in a bistable potential under the action of a nonconservative linear force field,
            \begin{align}\label{eq:Langevin2d}
                \dot x& = -4\alpha x^3 + 2 \beta x  - \gamma y +  \xi_x\\
                \dot y& = -2 \delta y +\gamma x +  \xi_y,
            \end{align}
        with $\sqrt{2\epsilon}(\xi_1,\xi_2)\equiv (\xi_x,\xi_y) $.
        We can solve the second equation in the stationary state
            \begin{align}
                y(t)& = \int_{-\infty}^t dt' e^{-2 \delta (t-t')} (\gamma  x(t')-\xi_y(t'))
            \end{align}
        and plug it into \eqref{eq:Langevin2d} to obtain the integro-differential equation
            \begin{align}\label{eq:Langevin_Integral}
                \dot x& = -4\alpha x^3 + 2 \beta x  - \gamma^2  \int_{-\infty}^t dt' e^{-2 \delta (t-t')}   x(t')+ \eta(t)
            \end{align}
        with nonMarkovian Gaussian noise $\eta(t) \equiv \xi_x(t) - \gamma \int_{-\infty}^t dt'  e^{-2 \delta (t-t')} \xi_y(t')  $ . By differentiating \eqref{eq:Langevin_Integral} with respect to time we get
            \begin{align}\label{eq:dot_Langevin_Integral}
                \ddot x& = -12\alpha \dot x x^2  + 2 \beta \dot x  - \gamma^2  x(t)+ 2 \gamma^2 \delta \int_{-\infty}^t dt'  e^{-2 \delta (t-t')} x(t')  + \dot \eta(t).
            \end{align}
        The integral appearing in \eqref{eq:dot_Langevin_Integral} can be eliminated by means of \eqref{eq:Langevin_Integral}, i.e.,
            \begin{align}
                \gamma^2  \int_{-\infty}^t dt' e^{-2 \delta (t-t')}   x(t') & = -4\alpha x^3 + 2 \beta x -   \dot x + \eta(t),
            \end{align}
        to obtain the underdamped Langevin equation
            \begin{align}\label{eq:Under_dot_Langevin_eff}
                \ddot x& = -12\alpha \dot x x^2  + 2 \beta \dot x  - \gamma^2 x- 8\alpha\delta x^3 + 4 \beta \delta x -  2\delta \dot x + 2\delta\eta   + \dot \eta(t)= -\Gamma(x) \dot x - \mathcal{U}'(x)  +2\delta \eta(t)   + \dot \eta(t),
            \end{align}
        with effective friction coefficient $\Gamma(x)\equiv 12\alpha  x^2 + 2 (\beta - \delta)$ and effective potential $\mathcal{U}(x)=   2 \alpha \delta x^4 + (\frac{\gamma^2}{2}-2\beta \delta)x^2 $. The friction coefficient $\Gamma$ is negative for $|x|< \sqrt{\frac{\beta-\delta}{6\alpha}}$, implying that energy is pumped into the system. 
        The noise reads
            \begin{align}
                \chi :=2\delta \eta  + \dot \eta &= 2 \delta \xi_x - 2 \delta \gamma \int_{-\infty}^t dt'  e^{-2 \delta (t-t')} \xi_y(t')  + \dot \xi_x + 2 \delta  \gamma \int_{-\infty}^t dt'  e^{-2 \delta (t-t')} \xi_y(t') +\gamma \xi_y \\
                &=2 \delta \xi_x + \dot \xi_x + \gamma \xi_y,
            \end{align}
        with correlation function
            \begin{align}
                \mean{\chi(t) \chi(0)} &= 4 \delta^2 \mean{\xi_x(t)\xi_x(0)} + \gamma^2 \mean{\xi_y(t)\xi_y(0)} - \frac{d^2}{dt^2} \mean{\xi_x(t)\xi(0)} \\
                &=2 \epsilon[(4 \delta^2 + \gamma^2) \delta(t) - \frac{d^2}{dt^2} \delta(t)]
            \end{align}

        The noiseless limit of \eqref{eq:Under_dot_Langevin_eff} is a Lienard equation \cite{perko2013differential} and for $\gamma \gg 1$, since the quadratic part of the potential becomes dominant, reads
            \begin{align}\label{eq:dot_Langevin_eff}
                \ddot x& = - \gamma^2  x - \Gamma(x)\dot x ,
            \end{align}
        which maps exactly onto the van der Pol oscillator with weak damping after rescaling the space coordinate $x$ and the time $t'=\gamma t $:
            \begin{align}\label{eq:van_der_pol}
                \ddot x& = -   x - \frac{1}{\gamma} \dot x (x^2-1) .
            \end{align}
        The latter is known to have oscillations of constant amplitude and period at leading order in $\gamma \gg 1$ \cite{andersen1982power}. This implies that in the original model $\ell \sim \gamma^0$ for $\gamma \gg1$ and $t_\text{p} \sim \gamma^{-1}$.

        Concerning the scaling of the Floquet mode $\zeta$, the biorthogonality condition only imposes that $\zeta_\parallel \sim \gamma^{-1} $ while the component $\zeta_\perp$ is unconstrained. However, we can estimate the correlation time $\tau_c$ considering the weak noise version of \eqref{eq:van_der_pol}, linearized around the periodic motion. Namely, setting $w := (x- \mathcal{x})/\sqrt{\epsilon}$  , we obtain the linear Langevin equation
            \begin{align}\label{eq:van_der_pol_linearized}
                \ddot{ w}& = -  w - \frac{1}{\gamma} \dot{\mathcal{x}} 2 w \mathcal{x} - \frac{1}{\gamma} \dot {w} (\mathcal{x}^2-1)  +  \gamma^{-1/2}\tilde \chi,
            \end{align}
        with $\chi(t)/(\sqrt{\epsilon}\gamma^2) = \tilde \chi(t')/\sqrt{\gamma}$ having correlations $\mean{ \tilde \chi(t') \tilde \chi(0)}=2 (\delta(t')-\frac{d^2}{dt'^2} \delta(t'))$ since 
            \begin{align}
                \mean{\chi(t) \chi(0)}&=2 \epsilon[(4 \delta^2 + \gamma^2) \delta(t) - \frac{d^2}{dt^2} \delta(t)]=2 \epsilon [\gamma(4 \delta^2 + \gamma^2) \delta(t') - \gamma^3\frac{d^2}{dt'^2} \delta(t')]\\
                &=2 \epsilon \gamma^3 [\delta(t')-\frac{d^2}{dt'^2} \delta(t')] + O(\epsilon\gamma)=  \epsilon \gamma^3 \mean{ \tilde \chi(t') \tilde \chi(0)} + O(\epsilon\gamma).
            \end{align}
        The drift field displays a small time-dependent correction $\frac{d \mathcal{x}^2}{dt'} \frac 1 \gamma$ to the natural frequency of the oscillator, as well as a linear friction term with a small, time-dependent friction coefficient $(\mathcal{x}^2-1)/\gamma$. The Gaussian noise comes from the leading term in the noise of \eqref{eq:dot_Langevin_eff} expressed in the new time coordinate and scaled by the noise intensity. Since the damping is small, the $x$-autocorrelation function will decay slowly with respect to the oscillating function $\mathcal{x}(t')$. So, in order to estimate $\tau_c$ we can substitute the time-dependent functions in \eqref{eq:van_der_pol} with their time-averaged values. This is called the method of averaging and works well in the deterministic case to evaluate the limit-cycle amplitude \cite{sanders2007averaging}. In particular, we obtain an effective friction coefficient
            \begin{align}
                \frac{1}{\gamma} (\overline{\mathcal{x}^2}-1)= \frac{1}{\gamma} \left[ 1+O(\gamma^{-1}) \right],
            \end{align}
        where we used the known asymptotic solution of the van der Pol equation $\mathcal{x} =2 \cos(t) +O(\gamma^{-1})$ and $\overline{\cos^2(t)}=1/2$ \cite{sanders2007averaging}. The time-averaged correction to the natural oscillation frequency vanishes at leading order, $\overline{\cos(t)\sin(t)}=0$. 
        Then, since \eqref{eq:van_der_pol} is linear, the position autocorrelation function can be obtained by Fourier transform. It will be exponentially suppressed as
            \begin{align}
                \mean{w(t') w(0)  } \propto \exp \left( -\frac {|t'|}{\gamma} \left[ 1+O(\gamma^{-1}) \right] \right ) = \exp \left( -|t| \left[ 1+O(\gamma^{-1}) \right] \right )
            \end{align}
        which implies that $\tau_c \sim \gamma^0 $ at leading order.

    \subsection{\emph{Numerics: fitting asymptotic values}}

\begin{figure}[h]
                \includegraphics[width=0.5\linewidth]{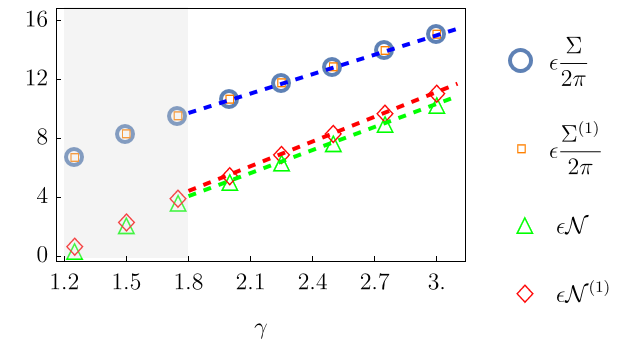}
                \caption{Numerics shows the linear growth for large $\gamma$. Linear fit obtained excluding points in the shaded region.}
                \label{fig: fitDW}
            \end{figure}

        To validate the previous analysis we have fitted the data in Fig. \ref{fig: hierarchyDW} in the Letter with a linear model $f(x)=A + Bx$, obtaining the set of parameters in table \ref{tab: fitDW}. Given that our predicted linear dependence in $\gamma$ is expected to occur at large forcing, we have dropped the first values at $\gamma<2$. 
            \begin{table}[h]
            \caption{Fitting parameters and coefficient of determination $R^2$ for $\Sigma/(2\pi), \Sigma^{(1)}/(2\pi),\mathcal{N}^{(1)}, \mathcal{N}$ in the stirred colloid example.}
                \centering
                \begin{tabular}{cccc}\hline
                 data  & A & B & $R^2$\\ \hline
                  $\Sigma/(2\pi), \Sigma^{(1)}/(2\pi)$ & $1.801\pm 0.008$ & $4.404\pm 0.003$ & $0.999998$\\
                  $\mathcal{N}^{(1)}$ & $-5.7 \pm 0.1$ & $5.61\pm 0.05$ & $0.999788$ \\
                  $\mathcal{N}$ & $-5.34 \pm 0.05$ & $5.23\pm 0.02$ & $0.99996$ \\
                  \hline
                \end{tabular}
                \label{tab: fitDW}
            \end{table}
        The goodness of our fits is certified by $R^2$ test of residuals, which provided values $\sim 1$ for all the quantities in our bounds. The interpolations are shown in Fig. \ref{fig: fitDW}. 

\begin{figure}[t]
                \centering
                \includegraphics[width=0.45\linewidth]{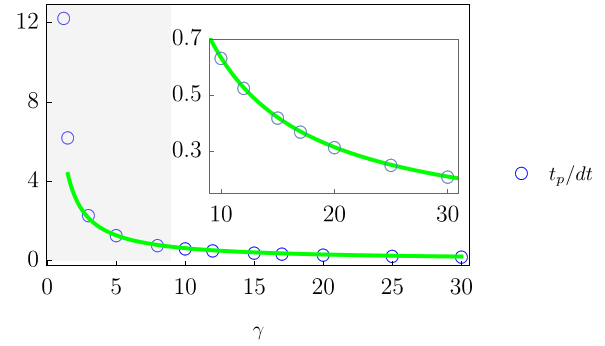}
                \includegraphics[width=0.45\linewidth]{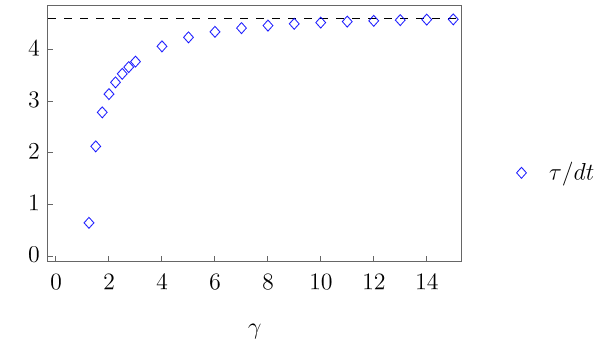}
                \caption{(Left) Nonlinear fit for the period of oscillations. Shaded area shows excluded points. Values obtained via integration along deterministic trajectory. Inset: magnification of the $1/\gamma$ behavior of fitted points. (Right) Asymptotic behavior of $\tau$, showing the predicted $\gamma^{0}$ trend.}
                \label{fig:tp_fit}
                \label{fig:tau_asymp}
            \end{figure}

        In addition, having stored the value of the noiseless period of oscillations $t_p$ at each value of $\gamma$, we can also show that asymptotically this quantity decreases as $t_p\sim\gamma^{-1}$. We observe that with the nonlinear model fit $y=A+Bx^n$, we obtain for the period the following set of parameters: $A = 6.71\pm 0.03$, $n= - 1.030\pm 0.003$, $B = 0.008\pm 0.001$. The interpolation is reported in Fig. \ref{fig:tp_fit} (left). The estimate improves as we get rid of more than just the first five elements of the dataset, but still shows the goodness of the asymptotic analysis.
            
        We conclude the far from equilibrium verifying that $\tau\sim \gamma^{0}$, as obtained in section \ref{sec VdP}. The numerical evidence of this asymptotic is reported in Fig. \ref{fig:tau_asymp} (right).
        
\section{Ring oscillator}

    \subsection{\emph{Supercritical Hopf bifurcation}}

        We linearize the deterministic dynamics of the ring oscillator close to the fixed point, $x^*=0$, and investigate the system local stability. Here, we focus on the example used to perform the numerics in the Letter, i.e. $N=3$.
      
        By expanding for $|x(t)-x^*|=|x(t)|$ small, we obtain the equation  $\dot{x} = J (x^*) \cdot x$, where $J(x^*)$ is the Jacobian matrix evaluated at the fixed point. The Jacobian matrix is defined as
            \begin{align}
                J(x_1,x_2,x_3) = \begin{pmatrix}
                    \partial_{x_1}F_1(x_3,x_1) & 0 & \partial_{x_3}F_1(x_3,x_1) \\
                    \partial_{x_1}F_2(x_1,x_2) & \partial_{x_2}F_2(x_1,x_2) & 0 \\
                    0 & \partial_{x_2}F_3(x_2,x_3) & \partial_{x_3}F_3(x_2,x_3)
                \end{pmatrix}\nonumber
            \end{align}

        with $F_i$ ($i=1,2,3$) the drift field given in the main text. Each $F_i$ depends only on two out of the three variables that define the system, so the zero in each row of the Jacobian matrix. 
        The eigenvalues of this matrix are
            \begin{align}
                \lambda_1 &= -\dfrac{2 \mathrm{e}^{V_{dd}/V_T}}{V_T}\\
                \lambda_2 &= \dfrac{(-3 + \mathrm{e}^{V_{dd}/V_T}) V_T^3 - \sqrt{3} \sqrt{-(-1 + \mathrm{e}^{V_{dd}/V_T})^2 V_T^6}}{V_T^4} \\
                \lambda_3 &= \dfrac{(-3 + \mathrm{e}^{V_{dd}/V_T}) V_T^3 + \sqrt{3} \sqrt{-(-1 + \mathrm{e}^{V_{dd}/V_T})^2 V_T^6}}{V_T^4}.
            \end{align}

        The first eigenvalue $\mu_1$ is real and negative $\forall V_{dd}$ bias, providing a stable direction for the dynamics. The other two conjugate eigenvalues $\mu_2,\mu_3$ are more interesting, as they are complex $\forall V_{dd}$, capturing the oscillating behavior of the device. A critical bias potential $V_{dd}^c = V_T\mathrm{log}3$ is found when the real part of $\mu_2,\mu_3$ vanishes.
        This is the signal of a supercritical Hopf bifurcation, with the fixed point that changes from being a stable spiral to a limit cycle in the three dimensional phase space after such critical point. 

    \subsection{\emph{Floquet theory for evaluation of} $\{\zeta^1\}$}

        The leading Floquet mode $\zeta^1$ (the superscript is dropped for sake of brevity in the main text) of the fundamental matrix of the linear dynamics of optimal fluctuations.
    
        The same approach has been also employed for the double-well, but we focus the discussion on the ring oscillator given the role played by the numerics to prove the bound far from the bifurcation. 
    
\begin{figure}[h!]
                \includegraphics[width=0.45\linewidth]{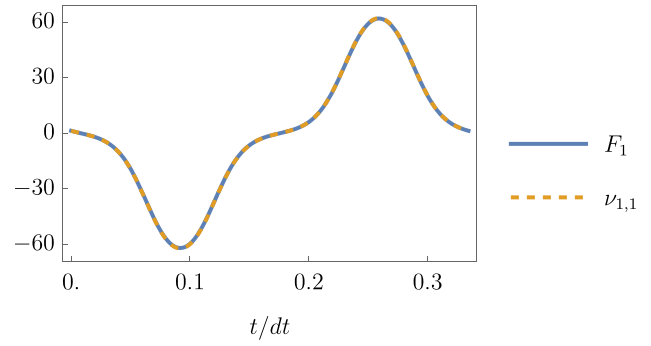}
                \includegraphics[width=0.45\linewidth]{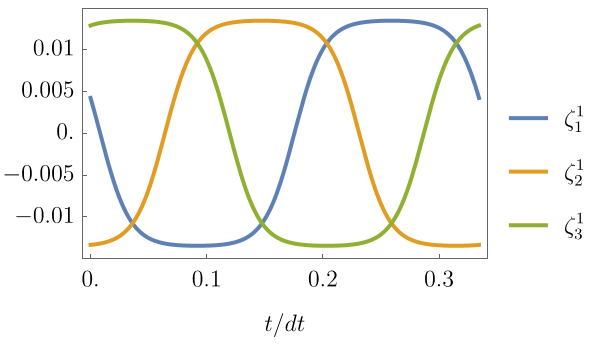}
                \includegraphics[width=0.45\linewidth]{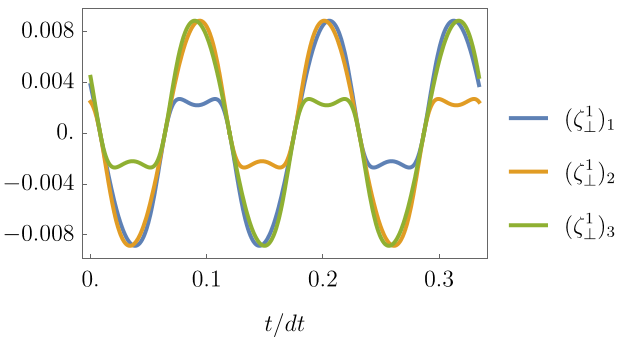}
                \includegraphics[width=0.45\linewidth]{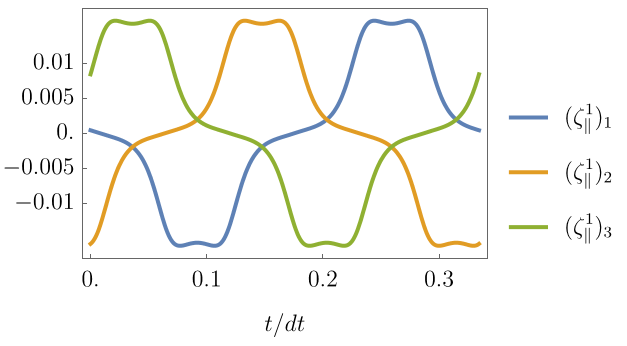}
                \caption{Numerical integration of the deterministic dynamics for $V_{dd}/V_{dd}^c = 2$. (Top left) The first component of the Floquet mode $\nu_1$ is shown to equal the first component of the drift  $F$ evaluated along the attractor. (Top right) Components of the Floquet mode $\zeta_1 \equiv \zeta$. (Bottom left)  Components of the perpendicular part $\zeta_\perp^1 \equiv \zeta_\perp$ of the Floquet mode $\zeta$. (Bottom right) Components of the tangent part $\zeta_\parallel^1 \equiv \zeta_\parallel$ of the Floquet mode $\zeta$.}
                \label{fig: EigFloquet}
            \end{figure}

        The strategy utilized can be summarized into the following algorithm \cite{vance1996fluctuations,gaspard2002trace}:
            \begin{enumerate}
                \item Having obtained the Jacobian $J$ after linearization of the system \mainref{Langevin} at $\epsilon=0$ in the main text,  we deal with an equation of the kind $\dot X(t) = J(t)\cdot X(t)$, which is the adjoint of the equation \mainref{linear}.
                \item We solve $\dot \Psi= J \cdot \Psi$ with $\Psi(0)=\mathbb{I}$ for the principal fundamental matrix $\Psi(t)$, i.e. the matrix whose columns are (normalized) linearly independent solutions of $\dot X(t) = J(t)\cdot X(t)$.
                \item We calculate the monodromy matrix $B := \Psi(t_p)$.
                \item We calculate the eigenvalues $\{\rho_j\}_{j=1...d}$ of $B$, know as  Floquet multipliers. We have that $\rho_1=1$ is zero up to numerical errors, as it correspond to the periodic solution of $\dot X(t) = J(t)\cdot X(t)$, and $\rho_{j \neq 1}<0$ since the periodic solution is stable.
                \item We calculate the eigenvectors $\{\xi_j\}_{j=1...n}$ of the monodromy matrix $B$. 
                \item From the Floquet multipliers $\{\rho_j\}_{j=1...n}$ we obtain the Floquet exponents $\{\mu_j\}_{j=1...d}$, defined as $\mu_j = \frac{1}{t_p}\ln \rho_j$. We have that $\mu_1=0$ is zero up to numerical errors.
                \item We obtain the Floquet modes $\{\nu_j\}_{j=1...d}$ which allow one to write the general solution of $\dot X(t) = J(t)\cdot X(t)$ as $X(t)=\sum_{j=1}^d a_j \nu_j e^{\mu_j t}  $. In particular, $\nu_1(t)=F(\mathcal{x}^*(t))$ corresponds to the deterministic drift field on the limit cycle, see Fig. \ref{fig: EigFloquet}. The Floquet modes are calculated as 
                    \begin{equation}
                        \nu_i(t) = \Psi(t) \cdot \xi_j \mathrm{e}^{-\mu_i t}.
                    \end{equation}
                \item With knowledge of the Floquet modes $\{\nu_j\}_{j=1...d}$, we obtain the Floquet modes $\{\zeta^j\}_{j=1...d}$ of the adjoint equation $\dot X = - J^T \cdot X$,  exploiting the biorthonormality condition, i.e., $\nu_i \cdot \zeta^j=\delta_{ij}$. The Floquet mode $\zeta^1$, that we rename $\zeta \equiv \zeta^1$ in the main text, is the one corresponding to the Floquet exponent $\mu_1=0$. 
            \end{enumerate}

        An example of the resulting $\zeta^1$, obtained by numerically integrating the dynamics for a fixed $V_{dd}=2$ is provided in Fig. \ref{fig: EigFloquet} in the top right panel. In the same figure we can observe the Floquet modes splitted into the parallel and orthogonal contributions such that $\zeta^1 = \zeta_{\parallel}^1+\zeta_{\perp}^1$, reported in components, in the bottom panels. It is evident the delay time $\Delta t$ between different components.

        \subsection{\emph{Numerical evaluation of} $\mathcal{I}$}

            \begin{figure}[htb]
                \centering
                \includegraphics[scale=0.9]{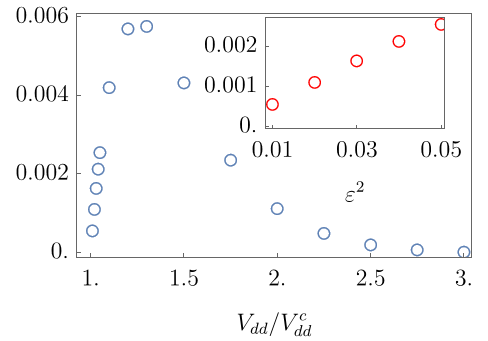}
                \caption{$\mathcal{I}$ as a function of the driving voltage $V_{dd}$ normalized to the bifurcation value $V_{dd}^c$. It is always nonnegative as announced in main text. Inset: the quantity approaches a finite value at the bifurcation with a linear trend in $\varepsilon^2= V_{dd} -V_{dd}^c$ as analytically derived in the Letter.}
                \label{fig: Iextra}
            \end{figure}

         We show with numerical simulations (see Fig. \ref{fig: Iextra}) that 
             \begin{align}
                \mathcal{I} := \int_0^{t_p} dt\, (\zeta_\perp \cdot D \cdot \zeta_\perp + 2\zeta_\parallel \cdot D \cdot \zeta_\perp  ) \geq 0
             \end{align}
        is verified even for $V_{dd}/V_{dd}^c\gtrsim 1$ as we report in the Letter. 
        The integral $ \mathcal{I}$ grows linearly with $V_{dd}-V^c_{dd}=\varepsilon^2$ close to the bifurcation (see inset of Fig. \ref{fig: Iextra}), as shown in the Letter, and it remains always positive, showing a peak around $V_{dd}/V_{dd}^c\sim 1.3$. The decreasing behavior shown afterwards does not mean that inequality \eqref{bound_tau_c} is saturated; in fact, a crucial role is played in the prefactor by $t_p^{-3}$, whose decrease with $V_{dd}/V_{dd}^c$ makes the difference $\tau^{-1}-(\tau^{(1)})^{-1}$ increase. With the evidence that $\tau_c^{(1)}\geq\tau_c$ we conclude that $\mathcal{N}\geq\mathcal{N}^{(1)}$ also holds for the ring oscillator with $N=3$. 

\section{Generalization to non-diagonal diffusion matrices, close to supercritical Hopf bifurcations}

    For the general case of a system governed by Langevin equation \mainref{Langevin} with non-diagonal diffusion matrix, we can show the validity of bound Eq. \mainref {series_bounds} in the proximity of a supercritical Hopf bifurcation. We denote $x^*_c$ the fixed point of the noiseless dynamical system at the bifurcation. The idea is to make a linear change of variable $x\mapsto z$ which results in a Langevin dynamics with uncorrelated noise components. Thus, we reduce our problem to the setting treated in the main text. It remains to link the correlation time 

     We start by recalling that the diffusion matrix can be decomposed as $D=AA$, and so the symmetric matrix $A$ is the square root of $D$, namely $A=D^{1/2}$. This matrix inherits the properties of positivity and invertibility, together with the periodicity of $D$ when evaluated on the limit cycle. Also, we choose the square root matrix such that $A=A^T$. We use this matric to define the new variable $z(t)=A(t)^{-1}(x(t)-x^*_c)$, such that
        \begin{equation}
            x-x^*_c = Az \quad, \quad \dot x = \dot A z + A \dot z
        \end{equation}
    in which $x^*_c$ is the fixed point of the deterministic dynamics at the bifurcation. The Langevin equation reads
        \begin{equation} \label{eq zDynamics}
            \dot z = A^{-1}[F(Az +x^*_c) - \dot A z] + \sqrt{2\epsilon}\xi \equiv\mathcal{F}(z)+ \sqrt{2\epsilon}\xi, 
        \end{equation}
    where the noise term is now completely uncorrelated $\mean{\xi}=0$, $\mean{\xi_i(t)\xi_j(t')} = \delta_{ij}\delta(t-t')$, being $\delta_{ij}$ the Kronecker delta. Given the drift term $\mathcal{F}(z)$ we are able to obtain the entropy production rate along the cycle, in the limit $\epsilon\to 0$
        \begin{align}\label{epr_z}
            \epsilon \dot\Sigma^{(z)} &= \mathcal{F}(z)\dot z\bigg\rvert_{\mathcal{z}^*(t)} =  |\mathcal{F}(z)|^2\bigg\rvert_{\mathcal{z}^*(t)}=  |A^{-1}[F(Az+x^*_c)-\dot A z]|^2\bigg\rvert_{\mathcal{z}^*(t)} \\
            & \leq [F^T D^{-1} F + (\dot A A^{-1} (x-x^*_c))^T D^{-1} \dot A A^{-1} (x-x^*_c)]\big|_{\mathcal{x}^*(t)} \\
            & = [\epsilon \dot\Sigma+ (\dot A A^{-1} (x-x^*_c))^T D^{-1} \dot A A^{-1} (x-x^*_c)]\big|_{\mathcal{x}^*(t)} \label{epr_z_unfolded}
        \end{align}
    having evaluated this quantity along the periodic orbit of the attractor, $ \mathcal{z}^*(t)=A^{-1}( \mathcal{x}^*(t)-x^*_c)$. Note that \eqref{epr_z} is not the entropy production rate of the coordinate $x$ subject to the mapping $x \mapsto z=A^{-1}(x-x^*_c)$. 
    In fact, \eqref{epr_z} is obtained regarding the field $\mathcal{F}(z)$ as even under time reversal, i.e., without swapping the sign of the term $\dot A$. 

    Given that the diffusion matrix in Eq. \eqref{eq zDynamics} is proportional to the identity, we conclude that the number of coherent cycles $ \mathcal{N}^{(z)}$ of the new coordinate $z$ fulfills the equivalent of \mainref{bound_N1}, namely,
        \begin{equation}
        \label{eq z_bound}
            \mathcal{N}^{(z)} = 2\pi\dfrac{\tau_c^{(z)}}{t_p^{(z)}} \leq \dfrac{1}{2\pi} \int_0^{t_p^{(z)}} dt\,\dot\Sigma^{(z)}.
        \end{equation}
    Note that, since $z(t)$ is a function of the $t_p$-periodic function $\mathcal{x}^*(t)$, $t_p$ is also a period of $\mathcal{z}(t)=\mathcal{z}(t+t_p)$, but not the minimum period, in general. Therefore, we set $t_p^{(z)}=t_p/n$ with integer $ n\geq 1$.

    Analogously to the procedure followed in the main text, we proceed weakening the bound \eqref{eq z_bound} in order to reach the full hierarchy involving corresponding quantities in terms of the original variable $x$, ruled by correlated noise. Let us start from the left-hand side; analyzing the autocorrelation of $x$, we observe that $\tau_c=\tau_c^{(z)}$. In fact, given the relation betwenn correlation functions
        \begin{equation}\label{corr_x_z}
            \mean{(x(t)-x^*_c)_i(x(0)-x^*_c)_i} = A_{ij}(t)A_{ik}(0)\mean{z_j(t)z_k(0)},
        \end{equation}
    and knowing that both $x$ and $z$ decorrelate exponentially as reported in the article
        \begin{align}
            \mean{x_i(t)x_i(0)} \sim \Theta(t) \mathrm{e}^{-t/\tau_c} \quad , \quad \mean{z_j(t)z_k(0)} \sim \Theta_z(t) \mathrm{e}^{-t/\tau_c^{(z)}},
        \end{align}
    with $\Theta(t)$ and $\Theta_z(t)$ periodic functions,
    we conclude that for the equality \eqref{corr_x_z} to hold true the condition $\tau_c=\tau_c^{(z)}$ must hold. Then, with this information about the decorrelation time and period, it is straightforward to write 
        \begin{equation}\label{eq NxNz}
            \mathcal{N}  = 2\pi\dfrac{\tau_c}{t_p} = 2\pi \dfrac{\tau_c^{(z)}}{t_p}n \leq 2\pi\dfrac{\tau^{(z)}}{t_p^{(z)}} =\mathcal{N}^{(z)}.
        \end{equation}
    To weaken the right-hand side of the bound \eqref{eq z_bound} we need to consider the kinetic term in \eqref{epr_z_unfolded}. If we consider a system that is close to the Hopf bifurcation, so as that the limit cycle is arbitrary small in the sense   $\max_t |\mathcal{x}^*(t)-x^*_c| = \varepsilon$ with $0 < \varepsilon \ll 1$. Therefore, the scaling $|F(\mathcal{x}^*(t))| \sim \varepsilon$ and $|\dot A(t)|=|\dot {\mathcal{x}}^*(t)\cdot  \nabla D^{1/2}(x^*_c) | \sim \varepsilon$ imply
        \begin{align}
            \label{eq SigmaExpanded}
               \dot\Sigma^{(z)}  
              & =  \dot\Sigma  + O(\varepsilon^4)
        \end{align}
    with $\dot\Sigma $ being of order $O(\varepsilon^2)$.    
    Thus, the right-hand side of \eqref{eq z_bound} reads
        \begin{equation}\label{eq rhsExpaned}
            \mathcal{N}^{(z)} \leq \dfrac{1}{2\pi} \int_0^{t_p/n} dt\, \left(\dot\Sigma  +O(\varepsilon^4) \right) \leq \dfrac{1}{2\pi} \Sigma  +O(\varepsilon^4,t_p),
        \end{equation}
    where in the last passage we have weakened the bound by extending the integral of the nonnegative quantity $\dot\Sigma $ from $t_p/n$ to $t_p$. Collecting Eqs. \eqref{eq z_bound}, \eqref{eq NxNz} and \eqref{eq rhsExpaned}, we obtain in the limit $\varepsilon \to 0$ the series of inequalities 
        \begin{equation}\label{eq FinalBoundXZ}
            \mathcal{N}  \leq \mathcal{N}^{(z)} \leq \dfrac{\Sigma^{(z)}}{2\pi} \leq \dfrac{\Sigma}{2\pi}\leq \dfrac{\Sigma_{\text{ME}}}{2\pi}.
        \end{equation}

    We conclude this section by stressing that Eq. \eqref{eq FinalBoundXZ} is obtained with no assumption on the diffusion matrix $D$, but it is required that the limit cycle originates continuously at the bifurcation, such as in the Hopf bifurcation. In fact, for other mechanisms creating a limit cycle---for example the global bifurcation for the double well in the main text---we cannot rely on the hypothesis of infinitesimal cycle radius used in the expansion in Eq. \eqref{eq SigmaExpanded}.

\section{The Brusselator model}\label{sec BrusSupp}

\begin{figure}[htb]
            \centering
            \includegraphics[width=0.5\textwidth]{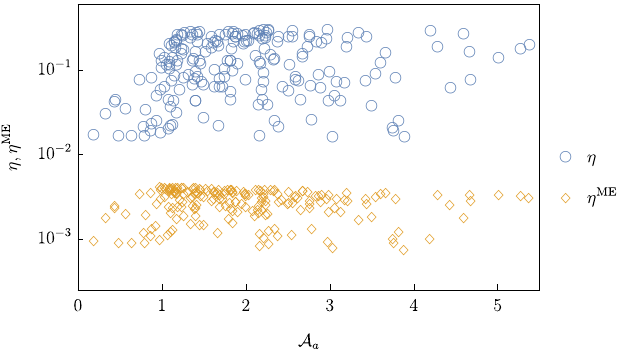}
            \caption{Numerical verification of Eq. \protect\mainref{bound_micro}, obtained via randomization of affinity $\mathcal{A}_a$.}
            \label{fig:LinAffBrus}
        \end{figure}

    We recall that the drift field $F_i= \sum_{\rho = \pm 4} \Delta_\rho^i j_\rho(x)$ entering the chemical Langevin equation for the macroscopic limit of the Brussaletor model reads 
        \begin{equation}
            \begin{pmatrix}
                F_1 \\
                F_2 
            \end{pmatrix}
            =
            \begin{pmatrix}
                 -k_{-3} x_1^3+k_3 x_1^2 x_2-k_{-1} x_1-k_2 x_1 y_2-k_4 x_1+k_{-2} x_2 y_3+k_{-4}y_4+k_1y_1 \\
                 k_{-3} x_1^3-k_3 x^2 x_2+k_2 x_1 y_2-k_{-2} x_2 y_3
            \end{pmatrix}, 
        \end{equation}
    and the associated nondiagonal diffusion matrix, whose elements are defined $D_{ij}(t) = \sum_{\rho=\pm 4}\Delta_{\rho}^i\Delta_{\rho}^j j_{\rho}(\mathcal{x}(t))/2$, is
        \begin{equation}\small
            D = \dfrac{1}{2}
                \begin{pmatrix}
                    k_1y_1 + k_{-1} x_1 + k_2 x_1 y_2 + k_{-2} x_2 y_3 + k_3 x_1^2 x_2 + k_{-3} x_1^3 + k_4 x_1 + k_{-4}y_4 & - (k_2 x_1 y_2 + k_{-2} x_2 y_3 + k_3 x_1^2 x_2 + k_{-3} x_1^3) \\
                    - (k_2 x_1 y_2 + k_{-2} x_2 y_3 + k_3 x_1^2 x_2 + k_{-3} x_1^3) & k_2 x_1 y_2 + k_{-2} x_2 y_3 + k_3 x_1^2 x_2 + k_{-3} x_1^3
                \end{pmatrix}
        \end{equation}   
    where $x_1$ and $x_2$ are evaluated on the periodic solution.
    The macroscopic reaction fluxes $j_\rho$ and the stoichiometric coefficients $\Delta^i_\rho$ are reported in the main text. In the main text we show the
    tightness of the bound by plotting $\eta, \eta^\text{ME}$ as a function of the the thermodynamic force, called affinity, $\mathcal{A}_b$. Here, instead, we choose the reaction parameters $k_1 = k_{-1} =  k_3 = k_{-3} = k_4 = k_{-4} = 1$, $k_{2}y_2  =4.8$ and $k_{-2}y_3= 0.06$ and vary $\mathcal{A}_a$ uniformly sampling the concentrations of the chemical species $y_1,y_4$ in the intervals $y_4\in [10^{-3}, 0.45]$, $y_3 \in [0.43, 2.05]$ and retaining only the values that result in an oscillatory steady state, for a total amount of 200. The ratios $\eta$ and $\eta^{\text{ME}}$ are shown in Fig. \ref{fig:LinAffBrus}. The tightness of the bound is similar to the case of varying  $\mathcal{A}_b$, with the dynamical entropy production associated to the Langevin equation $\Sigma$ offering a sizably better constraint than the thermodynamic entropy production associated to the master equation $\Sigma^{\text{ME}}$, i.e., $\eta /\eta^{\text{ME}} \simeq 10^2$ for most systems.

\end{document}